\documentclass[10pt,aps,pre,amsmath,onecolumn,superscriptaddress]{revtex4-2}
\usepackage{graphicx}
\usepackage{color}
\usepackage[dvipsnames]{xcolor}
\usepackage[caption=false]{subfig}
\usepackage{hyperref}

\newcommand{\bea}{\begin{eqnarray}}
\newcommand{\eea}{\end{eqnarray}}
\newcommand{\be}{\begin{equation}}
\newcommand{\ee}{\end{equation}}

\begin{document}

\title{Generating discrete-time constrained random walks and L\'evy flights}
\author{Benjamin \surname{De Bruyne}}
\affiliation{LPTMS, CNRS, Univ. Paris-Sud, Universit\'e Paris-Saclay, 91405 Orsay, France}
\author{Satya N. \surname{Majumdar}}
\affiliation{LPTMS, CNRS, Univ. Paris-Sud, Universit\'e Paris-Saclay, 91405 Orsay, France}
\author{Gr\'egory \surname{Schehr}}
\affiliation{Sorbonne Universit\'e, Laboratoire de Physique Th\'eorique et Hautes Energies, CNRS UMR 7589, 4 Place Jussieu, 75252 Paris Cedex 05, France}
\date{\today}

\begin{abstract}
  We introduce a method to exactly generate \emph{bridge} trajectories for discrete-time random walks, with arbitrary jump distributions, that are constrained to initially start at the origin and return to the origin after a fixed time. The method is based on an effective jump distribution that implicitly accounts for the bridge constraint. It is illustrated on various jump distributions and is shown to be very efficient in practice. In addition, we show how to generalize the method to other types of constrained random walks such as generalized bridges, excursions, and meanders.  
\end{abstract}

\maketitle
\section{Introduction}
Brownian motion lies at the heart of numerous applications in science. In its simplest form, the evolution of a one-dimensional \emph{free Brownian motion} $x(t)$ is governed by the Langevin equation
\begin{align}
  \dot x(t) = \sqrt{2\,D}\,\eta(t)\,,\label{eq:bm}
\end{align} 
where $D$ is the diffusion coefficient and $\eta(t)$ is a Gaussian white noise with zero mean $\langle\eta(t)\rangle=0$ and a delta-correlator $\langle\eta(t)\eta(t')\rangle=\delta(t-t')$. Simulating a free Brownian motion is easy: one just discretizes the time with increments $\Delta t$ in the Langevin equation (\ref{eq:bm}), which gives 
\bea \label{eq:discrete_L}
x(t+\Delta t) = x(t) + \sqrt{2D} \, \eta(t) \, \Delta t \;.
\eea
One then draws independently, at each step, a jump length $ \sqrt{2D} \eta(t) \Delta t$ distributed as a Gaussian with zero mean and variance $2 D \Delta t$. This simple procedure however does not work when the Brownian motion is constrained. Examples of constrained Brownian motions are abundant. For instance, there have been several studies on Brownian bridges, Brownian excursions, Brownian meanders, reflected Brownian motion, etc \cite{Yor2000,Majumdar05Ein,MP2010,Dev2010,PY2018}. These constrained Brownian motions appear naturally in many applications, ranging from ecology to finance and statistics \cite{Giu,convex1,convex2,Shepp79,MB2008,CB2012,Kol1933,Smirnov48}. For example, in the ecological context, animals foraging for food typically start from their nest and come back to the nest at the end of a fixed period $t_f$ \cite{MN1992,BLDS2009}. If their motion is described by Brownian motion, the trajectory of such walk is called a ``bridge'', i.e., the walk is constrained to come back to the same starting point. A more general bridge configuration corresponds to the case where the final position at time $t_f$ is fixed, but not necessarily the same as the starting point $x(0)$. 

A natural question then arises: what is an efficient algorithm to generate numerically constrained trajectories with the correct statistical weight? This is part of a more general question: how to efficiently sample atypical rare trajectories with a given statistical weight, which is typically very small? In general, rare trajectories are important as they capture specific information about the system that cannot be seen in the typical trajectories where observables concentrate around their mean. For instance, in the context of glasses, rare trajectories are key to understand the slow structural relaxation dynamics close to the glass transition where fluctuations are important \cite{Gar2018}. Because the analytical study of rare trajectories is usually intractable in practice, numerical methods to sample them are of primary interest and several methods have been developed for both equilibrium and out of equilibrium systems \cite{BCDG2002,GKP2006,GKLT2011,KGGW2018,Causer21}. Recently, reinforcement learning approaches were devised and have shown to be particularly efficient \cite{Rose21,Das21,Oakes20}.

  In the context of constrained Brownian motions that we discuss in this paper, a simple example of such a rare trajectory is a bridge where the walker comes back to the starting point after a fixed time $t_f$. The probability of such a trajectory is small, as it decays as $t_f^{-1/2}$ for large $t_f$. How do we generate such a Brownian bridge? A naive algorithm would be to generate all possible free Brownian trajectories $x(t)$ of duration $t_f$, starting at $x(0)=0$ and retain only those that come back to the close vicinity of the origin at $t=t_f$. Such a naive algorithm is of course rather wasteful and inefficient. Fortunately, for the one-dimensional Brownian bridge, a simpler way to generate a trajectory is to consider the process $X(t)$  
\begin{align}
 X(t) = x(t) - \frac{t}{t_f}\,x(t_f)\,,\quad t\in [0,t_f]\,,\label{eq:bcbm}
\end{align}
where $x(t)$ is a free Brownian motion starting at the origin $x(0)=0$. Note that the bridge condition $X(t_f) = X(0)=0$ is manifestly satisfied by the construction (\ref{eq:bcbm}). It is easy to show that the trajectories $X(t)$, generated via Eq.~(\ref{eq:bcbm}), have the correct statistical weight for a Brownian bridge, which is simply a Gaussian process. However, this construction is very specific for the continuous-time Brownian bridge and cannot be easily generalized to generate other constrained Brownian motions. It would then be nice to have a general method to generate constrained Brownian motions that are not specific to a particular type of constraints. Indeed, in probability theory, there exists the well known Doob transforms \cite{Doob,Pitman} that provide a prescription to construct constrained Markov trajectories \cite{CT2013,Rose21}. However, this method is not always completely explicit. For numerical purposes, one would like to write explicitly an effective Langevin equation that generates constrained Brownian motions (such as the bridge) with the correct statistical weight \cite{Orland,CT2013}.

Recently, for continuous time Markov processes, such an explicit effective Langevin equation was derived such that it automatically takes into account the constraints by adding an effective force in the Langevin equation~\cite{MajumdarEff15,CT2013}. For instance, a Brownian bridge with $X(0)=X(t_f) =0$ is generated by the effective Langevin equation~\cite{MajumdarEff15,CT2013}  
\bea \label{bridge_eff}
\dot X(t) = -\frac{X(t)}{t_f-t} + \sqrt{2D}\,\eta(t) \;,
\eea
where $\eta(t)$, as before, is a Gaussian white noise with zero mean and which is delta-correlated. In Eq.~(\ref{bridge_eff}), the first term is an effective constraint-force that drives
the particle to the final position $x(t_f)=0$ at time $t_f$. This trajectory can then be easily generated by time-discretizing the effective Langevin equation (\ref{bridge_eff}) as in the case of the free Brownian motion in Eq.~(\ref{eq:discrete_L}). This construction of an effective Langevin equation is rather versatile and can be extended to other constrained continuous-time Markov processes \cite{CT2013,MajumdarEff15,Baldassarri2021}, such as Brownian excursions, Brownian meanders, Ornstein-Uhlenbeck bridges and more recently to non-intersecting Brownian motions \cite{Grela2021}.

\begin{figure}[t]
  \begin{center}
    \includegraphics[width=0.5\textwidth]{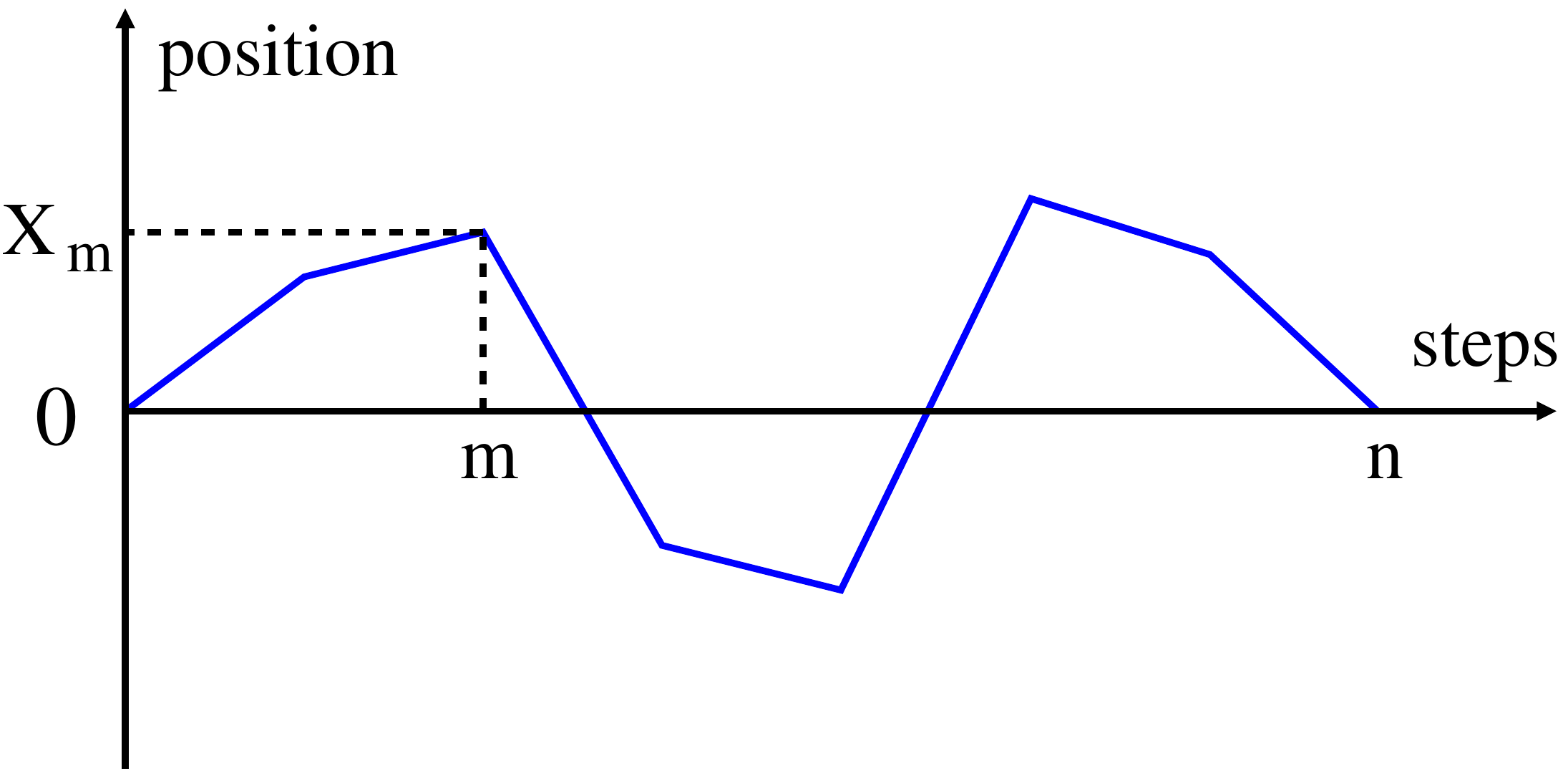}
    \caption{A bridge random walk of $n$ steps is a random walk that is constrained to start at the origin and return at the origin after $n$ steps. Due to the Markov property, a bridge random walk can be decomposed into two independent parts: a left part over the interval $[0,m]$, where it propagates from $0$ to $X_m$ and a right part over the interval $[m,n]$, where it propagates from $X_m$ to $0$.}
    \label{fig:bridge}
  \end{center}
\end{figure}

This construction of an effective Langevin equation works very nicely for continuous-time Markov processes. However, in many practical situations, stochastic processes are discrete in time and can be described by random walks evolving according to a discrete-time Markov rule 
\begin{align}
  x_{m+1} &= x_{m} + \eta_{m}\,, \label{eq:XmI}  
\end{align}
starting from $x_0 = 0$, where $\eta_m$'s are i.i.d.~random variables drawn from a distribution $f(\eta)$. Note that the jump distribution $f(\eta)$ may not have a finite second moment, such as in L\'evy flights where $f(\eta) \sim |\eta|^{-1-\mu}$ for large $|\eta|$, with $0<\mu<2$ denoting the L\'evy index. Generically, we refer to this discrete-time Markov jump processes as ``random walks''. Simulating a {\it free} discrete-time random walk (\ref{eq:XmI}) is rather straightforward, as in the discretized version of the free Brownian motion in Eq.~(\ref{eq:discrete_L}). However, like in the case of continuous-time Brownian motion, many natural phenomena are described by \emph{constrained} discrete-time random walks. For example, bridge random walks appear in many applications ranging from computer science to graph theory \cite{Majumdar05Ein,KnuthThe98,FlajoletOn98,MajumdarExact02,HararyDyna97,Takacs91,Schehr10Area,Perret12,Janson07}. Discrete-time bridge random walks also appear frequently in physics problems such as in fluctuating interfaces \cite{Ray,Antal,Majumdar04Flu1,Majumdar05Flu2,Schehr06SOS,MD2006,Gyo,Burk}, record statistics in time series \cite{GodrecheMecords15,GodrecheMecords17} or in anomalous diffusion of cold atoms \cite{Barkai1,Barkai2}. Furthermore, bridge random walks also play an important role in behavioral ecology \cite{Krebs93,Viswanathan11,GB2010} where the trajectories of foraging animals are tracked at discrete times using GPS. Moreover, the animals typically come back to their nests after a certain amount of time, which imposes the bridge constraint on the trajectories. In the context of mathematical finance, discrete-time bridge random walk models have been used to understand how Monte-Carlo  methods can be used to the valuation of mortgage backed security portfolios~\cite{Morokoff98}. A bridge random walk $X_m$ is a discrete-time process that evolves locally as in Eq.~(\ref{eq:XmI}), but is constrained to return to the origin after a fixed number of steps $n$ (see Fig.~\ref{fig:bridge}):
\begin{align}
    X_n = X_0=0\,.\label{eq:bridgecond}
\end{align}
 In many applications, it is often necessary to generate bridge trajectories numerically. Interestingly, generating bridge random walks is a challenging problem since a general prescription is not known for arbitrary jump distribution $f(\eta)$ in Eq.~(\ref{eq:XmI}). In the special case when the jump distribution is a pure Gaussian, i.e., $f(\eta) = e^{-\eta^2/2}/\sqrt{2 \pi}$, one can still generate bridge trajectories by using the discrete-time analogue of Eq.~(\ref{eq:bcbm}), namely
 \bea \label{bcbm_discr}
 X_m = x_m - \frac{m}{n} x_n \;.
 \eea
 However, this prescription does not work when $f(\eta)$ is not Gaussian. Another example where one can easily generate a bridge configuration corresponds to the $\pm 1$ random walk, where the jump distribution is $f(\eta) = (1/2) \delta(\eta+1) + (1/2) \delta(\eta - 1)$ \cite{Rose21,ChafaiRW12}. It is therefore important to develop an algorithm that does not depend on the specific form of the jump distribution. One possibility is to perform Markov chain Monte-Carlo simulations which consist in sampling the full joint jumps distribution $\{\eta_0\,,\ldots,\eta_{n-1}\}$ with the global bridge constraint that the total sum of the jumps is $\sum_{m=0}^{n-1} \eta_m=0$ \cite{Schehr10Area}. This Monte-Carlo method can also be computationally costly and require advanced techniques to probe the tails of distributions as the Monte-Carlo algorithm sometimes struggles to equilibrate the system. Given this absence of generic and efficient methods to generate constrained random walks, it is then highly desirable to derive an effective discrete-time jump process valid for arbitrary jump distributions $f(\eta)$, analogously to the effective Langevin equation in Eq.~(\ref{bridge_eff}) which is only valid for continuous-time Brownian bridge.
 
In this paper, we derive exactly an effective discrete-time jump process, valid for arbitrary jump distributions $f(\eta)$, to generate bridge random walks. As in the continuous-time Brownian bridge, we show that discrete-time bridges can be generated by an effective Markov jump process as in Eq.~(\ref{eq:XmI}), but the jumps $\eta_m$ have to be drawn from an effective distribution that depends on the bare jump distribution $f(\eta)$ and that effectively accounts for the bridge constraint. This general method is valid for arbitrary $f(\eta)$ and, as illustrations, we show how to compute explicitly the effective jump distribution in some examples. In certain cases, where the effective jump distribution is hard to sample directly, we provide an exact algorithm, based on the acceptance-rejection sampling (ARS) method, to sample these effective jump distributions (the code is available as a Python notebook in \cite{github}). Finally, we show that our method can be extended to other types of constrained discrete-time random walks, such as the ``generalized bridge'', the excursion, and the meander.  

The rest of the paper is organized as follows.  In Sec.~\ref{sec:bridgeRW}, we outline the derivation of the effective jump distribution for bridge random walks with an arbitrary jump distribution $f(\eta)$ and discuss the ARS method to draw samples from this effective jump distribution. In Sec.~\ref{sec:examples}, we discuss several examples to illustrate how the method works in practice. In Sec.~\ref{sec:gen} we generalize our method to other types of constrained processes such as the generalized bridge, the excursion, and the meander. Finally, we conclude in Sec.~\ref{sec:ccl} with a summary and perspectives. 

\section{Generating bridges for discrete random walks}
\label{sec:bridgeRW}

We start with the discrete-time random walk process $x_m$ in Eq.~(\ref{eq:XmI}) with an arbitrary jump distribution $f(\eta)$. We then show how
to construct a random walk bridge process $X_m$ of duration $n$ steps, satisfying the bridge constraint $X_n\!=\!X_0\!=\!0$. In Sec.~\ref{subsec1}, we show
how to generate $X_m$ via a Markov jump process with an effective jump distribution, valid for arbitrary $f(\eta)$. Subsequently, in Sec.~\ref{sec:ars}, we discuss
a general practical algorithm, based on the ARS method, to draw a jump length from this effective jump distribution. 

\subsection{Effective jump length distribution}\label{subsec1}

The derivation of the effective jump process for discrete-time random walk bridges follows closely the approach used
for the continuous time Brownian bridge developed in Ref.~\cite{MajumdarEff15}. The main idea is to first compute the probability
distribution $P_\text{bridge}(X,m\,|\,n)$ of the position of the walker for a random walk bridge at an intermediate time $m$, where $0 \leq m \leq n$.  
This can be easily computed using the Markov property of the free random walk process in Eq.~(\ref{eq:XmI}). Consequently, we derive an integral
equation for $P_\text{bridge}(X,m\,|\,n)$ from which we can read off the effective jump distribution.

Consider a bridge random walk trajectory in Fig.~\ref{fig:bridge} where the walk starts at the origin, returns to the origin after $n$ steps and arrives at $X$ at an intermediate time $m$. Using the Markov property, this trajectory can be decomposed into a left part over the interval $[0,m]$ and a right part over the interval $[m,n]$. Clearly the probability density $P_\text{bridge}(X,m\,|\,n)$ can then be written as
\begin{align}
    P_\text{bridge}(X,m\,|n) = {\cal N}\,{P(X,m|0,0)\,P(0,n|X,m)}\,, \label{eq:bridgePQ1}
 \end{align}
 where $P(X,m|X_0,m_0)$ is the forward propagator of a ``free'' random walk (without the bridge constraint) indicating the probability density to reach the position $X$ at step $m$, starting at $X_0$
 at step $m_0$ and ${\cal N}$ is a normalization constant which is fixed as follows. Integrating (\ref{eq:bridgePQ1}) over $X$ and setting it to $1$, as required by the normalization of the probability density, we get 
 \bea \label{eq:norm0}
 1 = {\cal N}\, \int_{-\infty}^{\infty}  {P(X,m|0,0)\,P(0,n|X,m)} \, dX  = {\cal N} P(0,n|0,0) \;,
 \eea
 where we have used the standard Chapman-Kolmogorov property of the transition probability (free propagator). This then gives
 \bea
  P_\text{bridge}(X,m\,|n) = \frac{P(X,m|0,0)\,P(0,n|X,m)}{P(0,n|0,0)}\,. \label{eq:bridgePQ0}
 \eea
The two terms in the numerator in Eq. (\ref{eq:bridgePQ0}) refer respectively to the left and the right part of the trajectories around step $m$. We start with the left part
of the trajectory characterised by $P(X,m|0,0)$. This satisfies the forward Kolmogorov equation  
\begin{align}
  P(X,m|0,0) = \int_{-\infty}^\infty dY\, P(Y,m-1|0,0)\,f(X-Y)\,,\label{eq:P0}
\end{align}  
with the initial condition $P(X,0|0,0) = \delta(X)$. This Eq. (\ref{eq:P0}) is obtained by considering the transition from step $m-1$ (where the particle is at $Y$) to step $m$ (where the particle is at $X$) and then integrating over all
possible values of $Y$. For the right part, it is convenient to write the backward Kolmogorov equation
\begin{align} \label{eq:Q0}
P(0,n|X,m) = \int_{-\infty}^\infty dY \, P(0,n|Y,m+1) f(Y-X)\, \;, \; {\rm for} \quad n \geq m +1 \;,
\end{align}  
with the condition $P(0,n|X,n) = \delta(X)$. This Eq. (\ref{eq:Q0}) is obtained by considering the transition from the initial step $m$ (where the particle is at $X$) to step $m+1$ (where the particle arrives at $Y$) and then integrating over all
possible values of $Y$. Note that, unlike in the forward Kolmogorov case, where one varies the ``final position'', in the backward Kolmogorov, one varies the initial position. Note that both equations (\ref{eq:P0}) and (\ref{eq:Q0}) are valid even when the jump distribution $f(\eta)$ is non-symmetric.     

Using the time translational invariance of the free propagator, these two Eqs. (\ref{eq:P0}) and (\ref{eq:Q0}) can be simplified further. For this, we 
will use the following shorthand notations (here and in the rest of the paper) 
 \bea \label{notations}
 P(X,m|0,0) \equiv P(X,m) \quad, \quad  P(0,n|X,m) \equiv Q(X,n-m) \quad, \quad P(0,n|0,0) \equiv P(0,n) \;,
 \eea
 where we have used the fact that the propagator of a free random walk depends only on the time difference between the final and the initial time. In terms of these shorthand notations (\ref{notations}), Eq. (\ref{eq:P0}) reads
 \begin{align}
  P(X,m) = \int_{-\infty}^\infty dY\, P(Y,m-1)\,f(X-Y)\,,\label{eq:P}
\end{align}
starting from the initial condition $P(X,0) = \delta(X)$. Similarly, Eq. (\ref{eq:Q0}) reads
\begin{align}
  Q(X,n-m) = \int_{-\infty}^\infty dY\, Q(Y,n-m-1)\,f(Y-X)\,, \; {\rm for} \quad n \geq m +1\;,   \label{eq:Q1}
\end{align}
starting from $Q(X,0) = \delta(X)$. Replacing $n-m$ by $m$ this can also be written in the standard backward form
\begin{align}
  Q(X,m) = \int_{-\infty}^\infty dY\, Q(Y,m-1)\,f(Y-X)\,, \; {\rm for} \quad m \geq 0 \;,   \label{eq:Q}
\end{align}
starting from the initial condition $Q(X,0) = \delta(X)$. Finally, in terms of these short-hand notations, Eq. (\ref{eq:bridgePQ0}) reads
\begin{align}
    P_\text{bridge}(X,m\,|n) = \frac{P(X,m)\,Q(X,n-m)}{P(0,n)}\,. \label{eq:bridgePQ}
 \end{align} 
Note that Eq.~(\ref{eq:bridgePQ}) has the following interpretation: out of all possible free paths that start from the origin and reach the origin at time $n$, the right hand side (rhs) of Eq.~(\ref{eq:bridgePQ}) counts the fraction of such paths that pass through the point $X$ at time $m$.

Our goal now is to write a forward Kolmogorov-type equation for the bridge propagator $P_\text{bridge}(X,m\,|n)$. For this purpose, we start with Eq.~(\ref{eq:bridgePQ}). Replacing $P(X,m)$ in the rhs of Eq.~(\ref{eq:bridgePQ}) by the rhs of Eq.~(\ref{eq:P}) we get
\bea \label{eq_eff_1}
P_\text{bridge}(X,m\,|\,n) = \int_{-\infty}^\infty dY\, \frac{P(Y,m-1)\, Q(X,n-m)}{P(X=0,n)}\, f(X-Y) \;.
\eea
Next we write Eq.~(\ref{eq:bridgePQ}) at time $m-1$ upon replacing $X$ by $Y$. This reads
\bea \label{eq_eff_2}
P_\text{bridge}(Y,m-1\,|\,n) = \frac{P(Y,m-1)\,Q(Y,n-m+1)}{P(X=0,n)} \;.
\eea
We then replace the ratio $P(Y,m-1)/P(X=0,n)$ on the rhs in Eq.~(\ref{eq_eff_1}) by its expression in Eq.~(\ref{eq_eff_2}). This gives us an integral equation for $P_{\rm bridge}(X,m|n)$
\begin{align}
   P_\text{bridge}(X,m\,|\,n)  &=  \int_{-\infty}^\infty dY \,P_\text{bridge}(Y,m-1\,|\,n)\, \tilde f(X-Y\,|\,Y,m-1,n)\,,\label{eq:effInt}
\end{align}
where the effective jump distribution $\tilde f(X-Y\,|\,Y,m,n)$ at time $m$ of the bridge of length $n$ is given by 
\begin{align}
  \tilde f(\eta\,|\,Y,m,n) = f(\eta) \,\frac{Q(Y+\eta,n-m-1)}{Q(Y,n-m)}\,. \label{eq:eff}
\end{align}
The effective distribution is therefore the free distribution that is modified in such a way that steps that take the walker closer to its final destination (here the origin) are more likely to happen. Note that this effective distribution is parametrized by the current position $Y$ of the bridge and moreover is non-stationary, i.e., it depends on the current time $m$ and also the total duration $n$. By using the equation for the backward propagator (\ref{eq:Q}), we can check that the effective distribution $\tilde f(\eta\,|\,Y,m,n)$ is normalized to unity, i.e., $\int_{-\infty}^\infty d\eta\,\tilde f(\eta\,|\,Y,m,n)=1$, irrespective of $Y, m$ and $n$. Note that Eq. (\ref{eq:eff}) can be viewed as an explicit representation of the generalized Doob transform which has been used previously \cite{CT2013,Rose21}.

Note that to evaluate the effective jump distribution $\tilde f(\eta\,|\,Y,m,n)$ in Eq.~(\ref{eq:eff}) where $Y$ is the position of the bridge at time $m$, we need to evaluate the backward propagator $Q(x,m)$ of the free walk. This can be explicitly computed from Eq.~(\ref{eq:Q}) by taking the Fourier transform with respect to $x$ and using the initial condition $Q(x,0) = \delta(x)$. This gives the backward propagator
\begin{align}
   Q(x,m) = \int_{-\infty}^\infty \frac{dk}{2 \pi}\,\left[\hat f(k)\right]^m\,e^{i\,k\,x}\;,\label{eq:QI}
\end{align}
where $\hat f(k)$ is the Fourier transform of the jump distribution $f(\eta)$, i.e.,
\bea \label{Fourier}
\hat f(k) = \int_{-\infty}^\infty d\eta \, f(\eta)\,e^{i k\eta} \;.
\eea
For certain specific jump distributions $f(\eta)$, it is possible to compute $\tilde f(\eta\,|\,Y,m,n)$ explicitly and sample directly from it -- we give several examples in Sec.~\ref{sec:examples}. For such cases where one has an explicit expression for $\tilde f(\eta\,|\,Y,m,n)$ along with a direct sampling method, one can easily draw a random number $\eta$ from this distribution to generate the bridge configuration numerically. However, in some cases, where such an explicit expression is difficult to obtain or when no direct sampling methods exist, how does one generate a random number from this distribution? In the next subsection, we show that in most cases, one can use an efficient and powerful numerical method to generate a random number distributed via $\tilde f(\eta\,|\,Y,m,n)$. 

\subsection{Acceptance-rejection sampling (ARS)}
\label{sec:ars}

\begin{figure}[t]
  \begin{center}
    \includegraphics[width=0.7\textwidth]{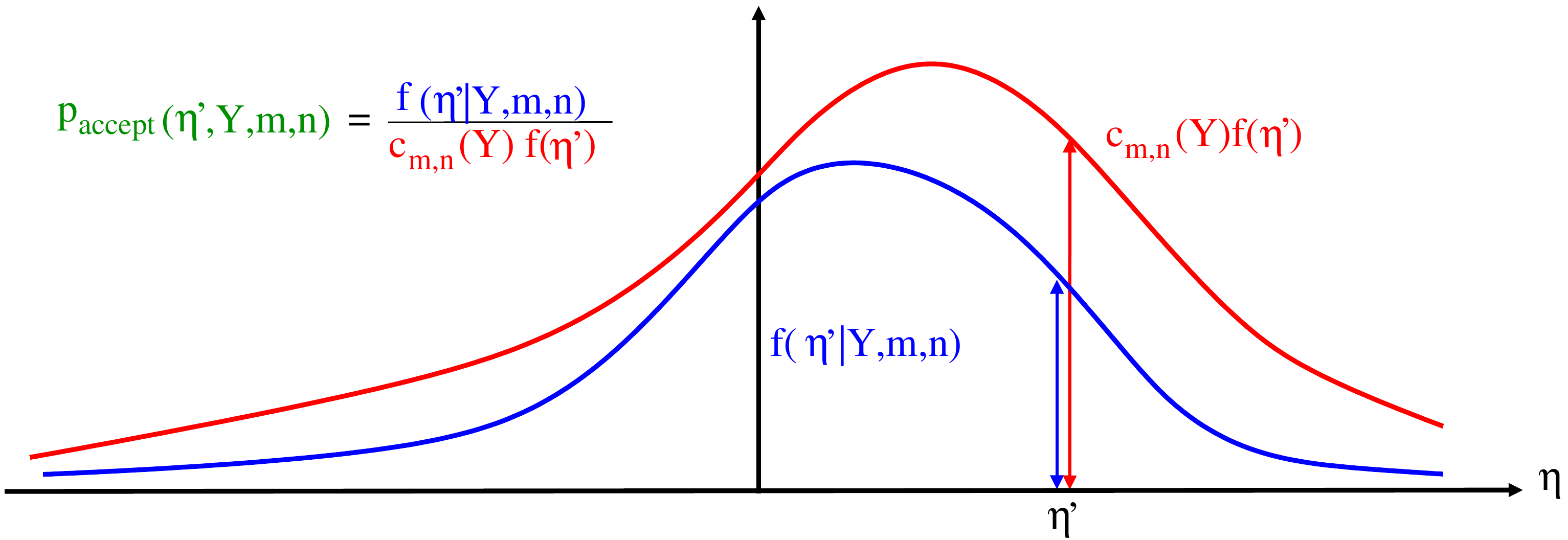}
    \caption{Illustration of the ARS method. To draw a sample from the effective jump distribution $\tilde f(\eta\,|\,Y,m,n)$ (blue curve), we draw a sample $\eta'$ from the free distribution $f(\eta')$ (red curve) and accept it with probability $p_\text{accept}(\eta',Y,m,n)=\frac{\tilde f(\eta'\,|\,Y,m,n)}{c_{m,n}(Y)\,f(\eta')}$ where $c_{m,n}(Y)$ is a constant such that the effective distribution (blue curve) is uniformly bounded by the free distribution (red curve). If the candidate is not accepted, we try again and look for another candidate.}
    \label{fig:acceptanceRejection}
  \end{center}
\end{figure}

Consider the effective jump distribution $\tilde f(\eta\,|\,Y,m,n)$ parametrized by $Y, m$ and $n$. We would like to draw a random number $\eta$
from this distribution. In this section, we discuss a powerful method that can generate such a random number efficiently. This method is based 
on ``acceptance-rejection sampling'' (see e.g. \cite{Gilks92}). For this method to work, we need one crucial condition, namely that 
\begin{align}
  \tilde f(\eta\,|\,Y,m,n) \leq c_{m,n}(Y)\,f(\eta) \,,\quad \forall \; \eta\;,\label{eq:ineq} 
\end{align}
where $c_{m,n}(Y) \geq 1$ is independent of $\eta$, such that the effective distribution $\tilde f(\eta\,|\,Y,m,n)$ is uniformly bounded by the free distribution $f(\eta)$ (see Fig.~\ref{fig:acceptanceRejection}). Assuming this condition is satisfied and we can estimate this constant $c_{m,n}(Y)$ explicitly, the algorithm proceeds in three steps:
\begin{enumerate}
  \item Draw a candidate random number $\eta'$ from the free distribution $f(\eta')$,
  \item Accept the candidate $\eta'$ with probability $p_\text{accept}(\eta',Y,m,n)$ given by 
  \begin{align}
    p_\text{accept}(\eta',Y,m,n)=\frac{\tilde f(\eta'\,|\,Y,m,n)}{c_{m,n}(Y)\,f(\eta')}\,,\label{eq:paccept}
  \end{align}
   \item Reject the candidate otherwise and look for another candidate from step 1.
\end{enumerate}
Note that $p_\text{accept}(\eta',Y,m,n)$ is just a number between $0$ and $1$ which depends on $\eta'$, $Y$, $m$ and $n$, but it should not be interpreted as a probability density function of $\eta'$. 
This algorithm indeed generates a random number from the distribution $\tilde f(\eta\,|\,Y,m,n)$. To show this, let us first compute the cumulative distribution of the accepted candidate. Our goal would be to show that this coincides with the target cumulative distribution, namely, $ \int_{-\infty}^{\eta}d\eta' \tilde f(\eta'\,|\,Y,m,n)$.
For the candidate random number $\eta'$ to be less than $\eta$, it must first be accepted which occurs with probability $p_\text{accept}(\eta',Y,m,n)$ and moreover its value must be less than $\eta$. Hence
\begin{align}
  \tilde F(\eta\,|\,Y,m,n) =\frac{\int_{-\infty}^{\eta}d\eta' f(\eta')\,p_\text{accept}(\eta',Y,m,n)}{\int_{-\infty}^{\infty}d\eta'\, f(\eta')\,p_\text{accept}(\eta',Y,m,n)}\,,\label{eq:Fmnexa}
\end{align}
where the denominator ensures the normalization, i.e., $\lim_{\eta \to \infty} \tilde F(\eta\,|\,Y,m,n) = 1$. Inserting the acceptance probability from Eq.~(\ref{eq:paccept}), we find
\begin{align}
  \tilde F(\eta\,|\,Y,m,n) =\frac{\int_{-\infty}^{\eta}d\eta' \tilde f(\eta'\,|\,Y,m,n)}{\int_{-\infty}^{\infty}d\eta'\, \tilde f(\eta'\,|\,Y,m,n)}=\int_{-\infty}^{\eta}d\eta' \tilde f(\eta'\,|\,Y,m,n)\,,\label{eq:Fmnproof}
\end{align}
which thus coincides with the target cumulative distribution. 

Finally, we note that average acceptance probability of the random number $\eta'$ is simply given by
\begin{align}
 \int_{-\infty}^{\infty}d\eta'\, f(\eta')\,p_\text{accept}(\eta',Y,m,n) = \frac{1}{c_{m,n}(Y)}\,,\label{eq:cmnlow}
\end{align}
which follows from the definition of the acceptance probability in Eq.~(\ref{eq:paccept}) and the normalization of $ \tilde f(\eta'\,|\,Y,m,n)$, i.e., $\int_{-\infty}^\infty d\eta'  \tilde f(\eta'\,|\,Y,m,n) = 1$. Since the left hand side of this equation is always less than one, we necessarily need $c_{m,n}(Y) \geq 1$. However, to make the algorithm efficient, we should maximize the average acceptance probability, i.e., we should try to use the smallest $c_{m,n}(Y) \geq 1$ that satisfies the inequality in Eq.~(\ref{eq:ineq}). In the next Section, we 
show how this algorithm can be successfully used for various jump distributions~$f(\eta)$.

\section{Examples}
\label{sec:examples}
\subsection{Lattice random walk}
\label{sec:latrw}
\begin{figure}[t]
\subfloat{%
 \includegraphics[width=0.5\textwidth]{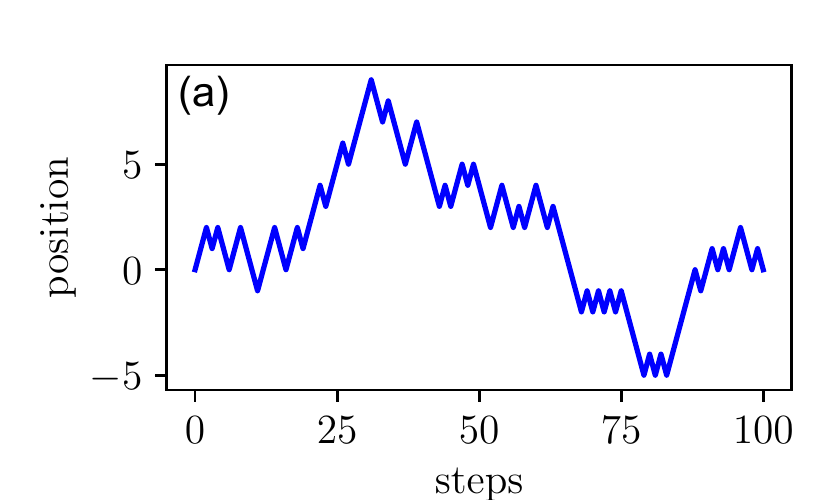}%
}\hfill
\subfloat{%
  \includegraphics[width=0.5\textwidth]{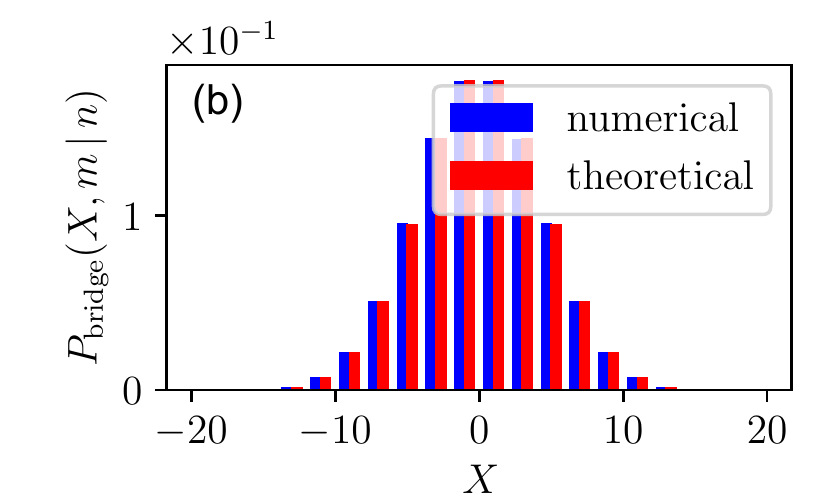}%
}\hfill
\caption{\textbf{Left panel (a):} A typical trajectory of a lattice bridge random walk of $n=100$ steps generated by the effective jump distribution in Eq.~(\ref{eq:effSim}). \textbf{Right panel (b):} Position distribution at $m=75$ for a lattice bridge of $n=100$ steps. The position distribution $P_{\rm bridge}(X,m|n)$ obtained numerically by sampling $10^5$ trajectories using the effective  distribution (\ref{eq:effSim}) is compared with the theoretical prediction in Eq.~(\ref{Pbridgelat}). }\label{fig:bridgeSim}
\end{figure}

We consider a lattice bridge random walk of $n$ steps. The free jump distribution is
\begin{align}
  f(\eta) = \frac{1}{2}\delta(\eta-1)+\frac{1}{2}\delta(\eta+1)\,.\label{eq:fSim}
\end{align}
The backward propagator $Q(y,m)$ in this case is well known and can be easily computed as follows. Let $n_+$ and $n_-$ denote the number of positive and negative jumps respectively that bring the walker from the initial position $Y$ to $0$ in $m$ steps. Clearly $n_+ + n _- = m$ and $n_+ - n_- = -Y$. Consequently $n_+ = (m-Y)/2$  and $n_- = (m+Y)/2$. Note that $Y$ has to be such that both $n_+$ and $n_-$ are integers. The probability that $n_-$ out of $m$ steps are negative is simply given by the binomial distribution $P(n_-|m) = \binom{m}{n_-} 2^{-m}$. Hence, replacing $n_-$ by $(m+Y)/2$ gives
the backward propagator
\begin{align}
  Q(Y,m) = \binom{m}{\frac{m+Y}{2}}\,2^{-m}\,, \label{eq:propsimple}
\end{align}
where $(m+Y)$ is even (otherwise $Q(Y,m)$ vanishes). We now substitute Eq.~(\ref{eq:propsimple}) in Eq.~(\ref{eq:eff}). First point to note is that, since $\eta = \pm 1$, the backward propagator appearing on both the numerator and the denominator on the rhs of  Eq.~(\ref{eq:eff}) are nonzero if and only if $Y+n-m$ is an even number. This condition is actually automatically satisfied for a lattice bridge, in the sense that for the walker to be at $Y$ at an intermediate time $m$ one must necessarily satisfy that $Y+n-m$ is even. Given this condition, the rhs of Eq.~(\ref{eq:eff}) can now be explicitly computed, giving
\begin{align}
   \tilde f(\eta\,|\,Y,m,n) = \frac{1}{2}\left(1-\frac{Y}{n-m}\right)\delta(\eta-1) + \frac{1}{2}\left(1+\frac{Y}{n-m}\right)\delta(\eta+1) \label{eq:effSim} \;.
  \end{align}
In this special case, it turns out that the ARS method is not needed and we can directly sample the jumps from the effective jump distribution in Eq.~(\ref{eq:effSim}) to generate bridge trajectories (see left panel in Fig.~\ref{fig:bridgeSim}). In the right panel in Fig.~\ref{fig:bridgeSim}, we computed numerically the probability distribution of the position at some intermediate time by generating bridge trajectories from Eq.~(\ref{eq:effSim}). This is compared to the theoretical position distribution for the bridge which can be easily computed by substituting the free propagators $P(X,m) = \binom{m}{\frac{X+m}{2} }2^{-m}$ and $Q(X,m)$ from Eq.~(\ref{eq:propsimple}) in Eq.~(\ref{eq:bridgePQ}), which gives
\bea \label{Pbridgelat}
P_{\rm bridge}(X,m|n) = \frac{\binom{m}{\frac{m+X}{2}} \binom{n-m}{\frac{n-m+X}{2}}}{\binom{n}{\frac{n}{2}}} \;,
\eea
which is nonzero only if $(m+X)$ as well as $n$ are both even numbers.  As can be seen in the right panel in Fig.~\ref{fig:bridgeSim}, the agreement is perfect.  
  
\subsection{Gaussian random walk}
\label{sec:Bg}
We consider a Gaussian bridge random walk of $n$ steps. In this case, the free jump distribution is 
\begin{align}
  f(\eta) = \frac{1}{\sqrt{2\pi\,\sigma^2}}\,e^{-\frac{\eta^2}{2\sigma^2}}\,.\label{eq:gaussDist}
\end{align} 
In this particular case, the bridge configuration can be generated by the explicit construction in Eq.~(\ref{bcbm_discr}) in terms of a free Gaussian walk. We show in 
Appendix \ref{app:gbridge} that this construction generates a Gaussian bridge with the correct statistical weight. It turns out that in this case one can also compute 
the effective jump distribution in Eq.~(\ref{eq:eff}) explicitly in terms of the backward propagator  
\begin{align}
  Q(Y,m) = \frac{1}{\sqrt{2\pi\,m\,\sigma^2}}\,e^{-\frac{Y^2}{2\,m\,\sigma^2}}\,.\label{eq:backPropG}
  \end{align}
This gives the effective jump distribution (\ref{eq:eff})   
\begin{align}
   \tilde f(\eta\,|\,Y,m,n) =\frac{1}{\sqrt{2\pi\,\sigma_{m,n}^2}}\,e^{-\frac{(\eta-\mu_{m,n})^2}{2\sigma_{m,n}^2}}\,,\label{eq:effG}
  \end{align}
  where
  \begin{align}
   \mu_{m,n}&=-\frac{Y}{n-m}\,,\label{eq:mumn}\\
   \sigma_{m,n}^2&=\frac{(n-m-1)\,\sigma^2}{n-m}\,.\label{eq:sigmamn}
\end{align}
Interestingly, the effective jump distribution is also a Gaussian but with a nonzero time-dependent mean $\mu_{m,n}$, while the bare jump
distribution $f(\eta)$ in Eq.~(\ref{eq:gaussDist}) is a Gaussian with {\it zero mean}. Thus the nonzero drift forces the trajectory to return to the
origin at time $n$. Note that the effective jump distribution (\ref{eq:effG}) is simply a rescaled and recentered Gaussian distribution. Therefore, the effective equation of motion for the bridge Gaussian walk can be written as
\begin{align}
  X_{m+1} = X_m - \frac{X_m}{n-m} + \frac{\sqrt{n-m-1}}{\sqrt{n-m}}\,\sigma\,\eta_m \quad, \quad 0 \leq m < n\,.\label{eq:eomG}
\end{align}
In the continuous time limit, this equation nicely converges to the effective Langevin equation in Eq.~(\ref{bridge_eff}). 

\begin{figure}[t]
\subfloat{%
 \includegraphics[width=0.5\textwidth]{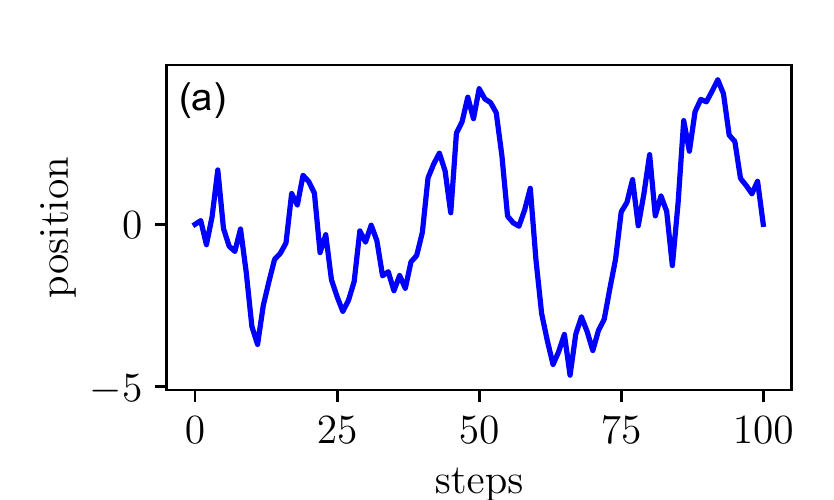}%
}\hfill
\subfloat{%
  \includegraphics[width=0.5\textwidth]{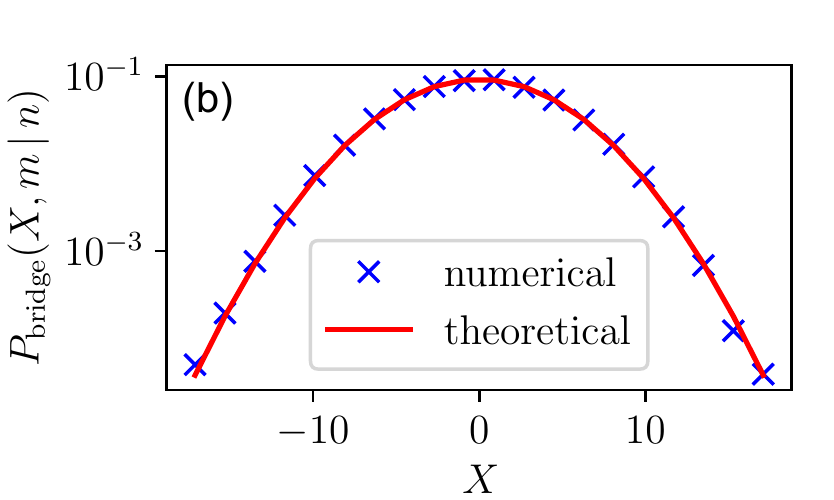}%
}\hfill\\
\subfloat{%
 \includegraphics[width=0.5\textwidth]{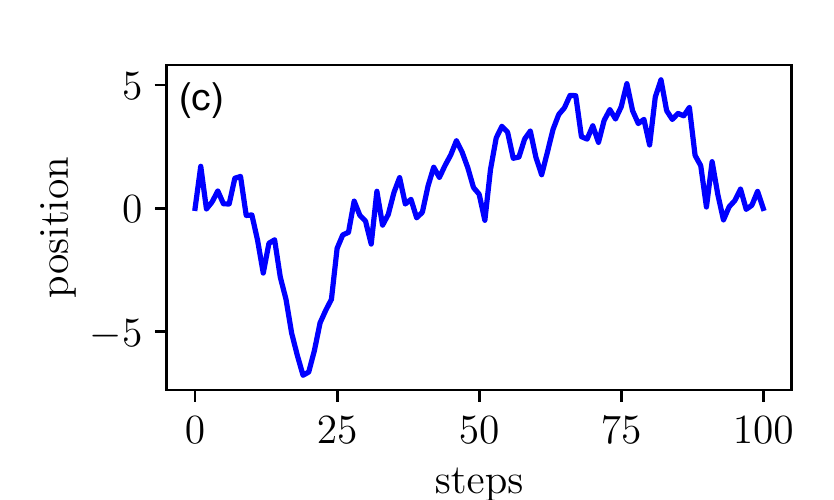}%
}\hfill
\subfloat{%
  \includegraphics[width=0.5\textwidth]{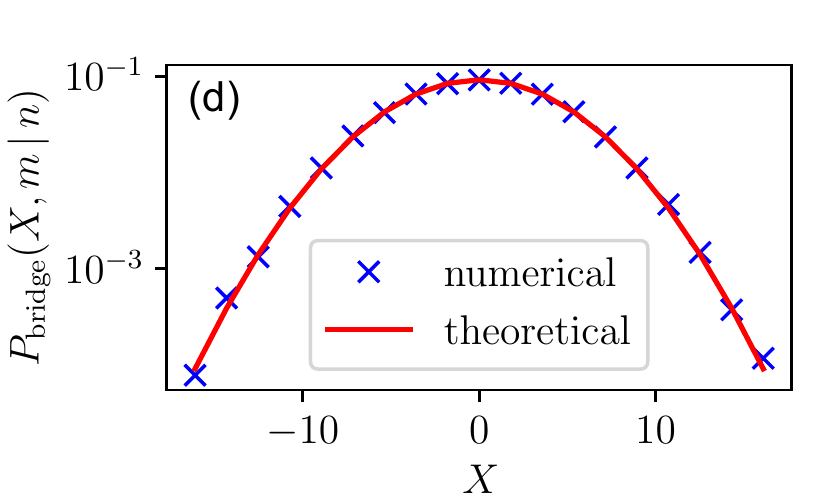}%
}\hfill\\
\caption{\textbf{Upper left panel (a):} Typical trajectories of a Gaussian bridge random walk of $n=100$ steps generated with a direct sampling of the effective distribution in Eq.~(\ref{eq:effG}). \textbf{Upper right panel (b):} Position distribution at $m=75$ for a Gaussian bridge of $n=100$ steps generated with a direct sampling by sampling $10^5$ trajectories using the effective distribution. The position distribution $P_{\rm bridge}(X,m|n)$ obtained numerically by sampling the jumps from the effective  distribution in Eq.~(\ref{eq:effG}) is compared with the theoretical prediction in Eq.~(\ref{PbridgeG}). \textbf{Lower left panel (c):} Typical trajectories of a Gaussian bridge random walk of $n=100$ steps generated using the ARS method on the the effective distribution in Eq.~(\ref{eq:effG}). \textbf{Lower right panel (d):} Position distribution at $m=75$ for a Gaussian bridge of $n=100$ steps generated using ARS on the the effective distribution. The position distribution $P_{\rm bridge}(X,m|n)$ obtained numerically by sampling $10^5$ trajectories using the ARS on the effective  distribution in Eq.~(\ref{eq:effG}) is compared with the theoretical prediction in Eq.~(\ref{PbridgeG}). }\label{fig:bridgeG}
\end{figure}

Finally, let us remark that for this Gaussian random walk model, we again do not need to use the 
ARS described in Sec.~\ref{sec:ars} since the effective jump distribution
in Eq.~(\ref{eq:effG}) is a Gaussian distribution from which one can sample bridge trajectories (see upper left panel in Fig.~\ref{fig:bridgeG}). Nevertheless, this jump distribution serves as a good illustration
on how the ARS method works. Let us recall that the main ingredient of the ARS method is to find
a constant (independent of the jump $\eta$) $c_{m,n}(Y) \geq 1$ such that Eq.~(\ref{eq:ineq}) is satisfied. To find this
constant, we consider Eq.~(\ref{eq:eff}). First we note the following inequality $Q(Y+\eta,n-m-1) \leq Q(0, n-m+1)$, for all $\eta$, which 
simply follows from the fact that $Q(X,n-m-1)$, as a function of $X$,  has a peak at $X=0$ since it has a Gaussian shape with zero mean.
Using this inequality in Eq.~(\ref{eq:eff}), we get, for $m<n$
\begin{align}
\tilde f(\eta\,|\,Y,m,n) \leq \frac{Q(0,n-m-1)}{Q(Y,n-m)}\,f(\eta)\,.\label{eq:fmnbound}
\end{align}
Comparing this to Eq.~(\ref{eq:ineq}), we see that a natural choice for $c_{m,n}(Y)$ is 
\begin{align}
  c_{m,n}(Y) = \frac{Q(0,n-m-1)}{Q(Y,n-m)} = \frac{\sqrt{n-m}}{\sqrt{n-m-1}}\,e^{\frac{Y^2}{2(n-m)\,\sigma^2}}\,,\label{eq:cG}
\end{align}
which manifestly satisfies $c_{m,n}(Y) \geq 1$ for all $m < n$ and all $Y$. This yields, from Eq.~(\ref{eq:paccept}), the following acceptance probability 
\begin{align}
  p_\text{accept}(\eta',Y,m,n) = e^{-\frac{(Y+\eta')^2}{2(n-m-1)}}\,.\label{eq:pacceptG}
\end{align}
One can thus use the ARS method, with this acceptance probability (\ref{eq:pacceptG}) to generate a Gaussian bridge (see lower left panel in Fig.~\ref{fig:bridgeG}). In the right panel in Fig.~\ref{fig:bridgeG}, we computed numerically the probability distribution of the position at some intermediate time by directly sampling the jump distribution from Eq.~(\ref{eq:effG}) (upper right) and by the ARS method with the acceptance probability in (\ref{eq:pacceptG}) (lower right). This is compared to the theoretical position distribution for the bridge which can be easily computed by substituting the free propagators $P(X,m) =\frac{1}{\sqrt{2\pi\,m\,\sigma^2}}\,e^{-\frac{X^2}{2\,m\,\sigma^2}}$ and $Q(X,m)$ from Eq.~(\ref{eq:backPropG}) in Eq.~(\ref{eq:bridgePQ}). Thus the theoretical position distribution is given by
\bea \label{PbridgeG}
P_{\rm bridge}(X,m|n) = \frac{1}{\sqrt{2\pi\sigma^2}}\,\sqrt{\frac{n}{m(n-m)}}\,e^{-\frac{n\,X^2}{2\,m(n-m)\sigma^2}} \;.
\eea
 As can be seen in Fig.~\ref{fig:bridgeG} (upper and lower right panel), the agreement is perfect.  

\subsection{Cauchy random walk}
\label{sec:cau}

Our next example is a Cauchy random walk bridge of $n$ steps. The normalized free jump distribution is symmetric with divergent moments
\begin{align}
  f(\eta) = \frac{1}{\gamma\,\pi}\,\frac{1}{\left[1+\left(\frac{\eta}{\gamma}\right)^2\right]}\,,\label{eq:cauchyDist}
\end{align} 
where $\gamma$ is a parameter that provides the typical scale of the jumps. Being a stable distribution, the backward propagator is simply (see e.g. \cite{BouchaudAn90})
\begin{align}
  Q(Y,m) = \frac{1}{\gamma\,m\,\pi }\frac{1}{\left[1+\left(\frac{Y}{\gamma\,m}\right)^2\right]}\,.\label{eq:cauchyProp}
  \end{align}
The effective step distribution at the $m^\text{th}$ step (\ref{eq:eff}) is therefore given by 
\begin{align}
  \tilde f(\eta\,|\,Y,m,n) = \frac{1}{\gamma\,\pi}\,\frac{n-m}{\left(n-m-1\right)}\,\frac{1}{\left[1+\left(\frac{\eta}{\gamma}\right)^2\right]}\,\frac{1+\left(\frac{Y}{\gamma(n-m)}\right)^2}{\left[1+\left(\frac{Y+\eta}{\gamma(n-m-1)}\right)^2\right]} \;.\label{eq:effC}
  \end{align}
Note that, unlike the free distribution $f(\eta)$ in Eq.~(\ref{eq:cauchyDist}), the effective distribution in Eq.~(\ref{eq:effC}) is asymmetric, has a power law tail 
$\tilde f(\eta\,|\,Y,m,n)\propto 1/\eta^4$ as $|\eta| \to \infty$ and consequently has a finite second moment. 

\begin{figure}[t]
\subfloat{%
 \includegraphics[width=0.5\textwidth]{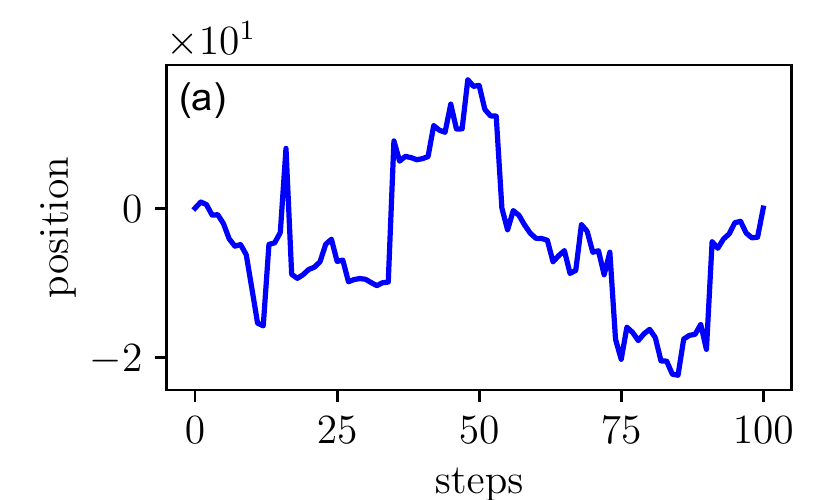}%
}\hfill
\subfloat{%
  \includegraphics[width=0.5\textwidth]{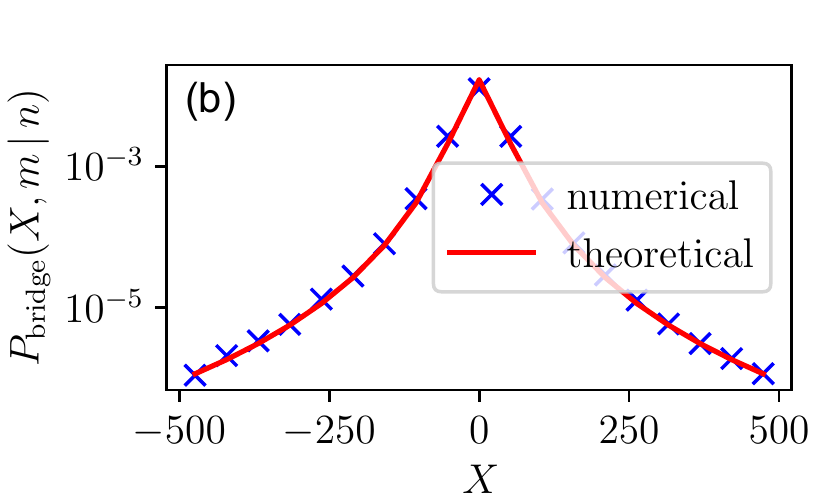}%
}\hfill
\caption{\textbf{Left panel (a):} Typical trajectories of a Cauchy bridge random walk of $n=100$ steps generated using the ARS method on the the effective distribution in Eq.~(\ref{eq:effC}). \textbf{Right panel (b):} Position distribution at $m=75$ for a Cauchy bridge of $n=100$ steps generated using ARS on the the effective distribution. The position distribution $P_{\rm bridge}(X,m|n)$ obtained numerically by sampling $10^5$ trajectories using the ARS on the effective  distribution in Eq.~(\ref{eq:effC}) is compared with the theoretical prediction in Eq.~(\ref{PbridgeC}). }\label{fig:bridgeC}
\end{figure}

Thus in this case, even though the effective jump distribution in Eq.~(\ref{eq:effC}) is explicit, it is not easy
to draw a random number from this distribution. Hence, this is the first example where the ARS method
discussed in Sec.~\ref{sec:ars} becomes handy. Again, for the ARS to work, we need to find a constant
$c_{m,n}(Y) \geq 1$, independent of $\eta$, that satisfies the inequality Eq.~(\ref{eq:ineq}). As in the Gaussian case, the propagator
$Q(Y,m)$ is peaked around $Y=0$ and hence, in Eq.~(\ref{eq:eff}) one can use the inequality $Q(Y+\eta,n-m-1) \leq Q(0,n-m-1)$.
Hence, using this inequality in Eq.~(\ref{eq:eff}), we see from Eq.~(\ref{eq:ineq}) that a natural choice for $c_{m,n}(Y) \geq 1$ is
\begin{align}
  c_{m,n}(Y) = \frac{Q(0,n-m-1)}{Q(Y,n-m)} = \frac{n-m}{n-m-1}\,\left[1+\left(\frac{Y}{\gamma(n-m)}\right)^2\right]\,.\label{eq:cauchycmn}
\end{align}
This yields the following acceptance probability, using Eq.~(\ref{eq:paccept}), 
\begin{align}
   p_\text{accept}(\eta,Y,m,n) = \frac{1}{1+\left(\frac{Y+\eta}{\gamma(n-m)}\right)^2}\,.\label{eq:cauchypa}
\end{align}
Using the ARS method discussed in Sec.~\ref{sec:bridgeRW}, we then generate the Cauchy bridge trajectories (see left panel in Fig.~\ref{fig:bridgeC}). In the right panel in Fig.~\ref{fig:bridgeC}, we computed numerically the probability distribution of the position at some intermediate time by sampling the jump distribution from Eq.~(\ref{eq:effC}) using the ARS method with the acceptance probability in (\ref{eq:cauchypa}). This is compared to the theoretical position distribution for the bridge that can be easily computed by substituting the free propagators $P(X,m) =\frac{1}{\gamma\,m\,\pi }\frac{1}{\left[1+\left(\frac{X}{\gamma\,m}\right)^2\right]}$ and $Q(X,m)$ from Eq.~(\ref{eq:cauchyProp}) in Eq.~(\ref{eq:bridgePQ}), which gives
\bea \label{PbridgeC}
P_{\rm bridge}(X,m|n) = \frac{1}{\gamma\,n}\,\frac{n}{m\,(n-m)}\,\frac{1}{1+\left(\frac{X}{\gamma\, m}\right)^2}\,\frac{1}{1+\left(\frac{X}{\gamma (n-m)}\right)^2} \;.
\eea
 As can be seen in Fig.~\ref{fig:bridgeC} (right panel), the agreement is perfect.  

\subsection{Student's random walk}
\label{sec:stud}
\begin{figure}[htb]
\subfloat{%
 \includegraphics[width=0.5\textwidth]{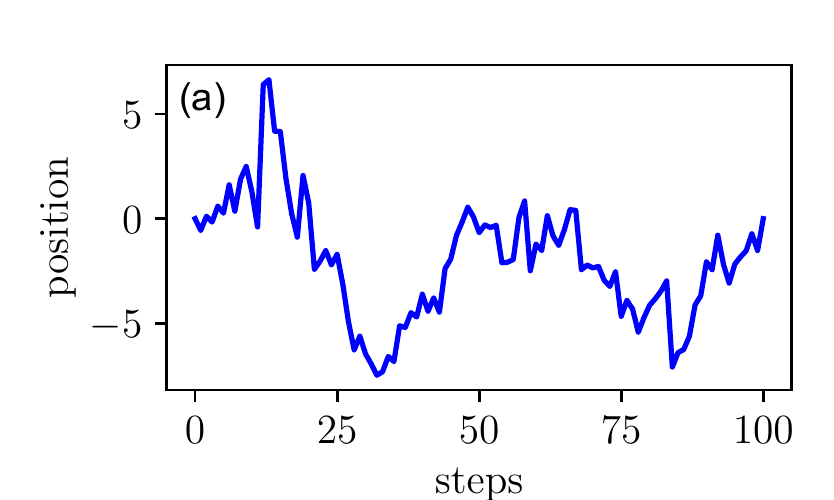}%
}\hfill
\subfloat{%
  \includegraphics[width=0.5\textwidth]{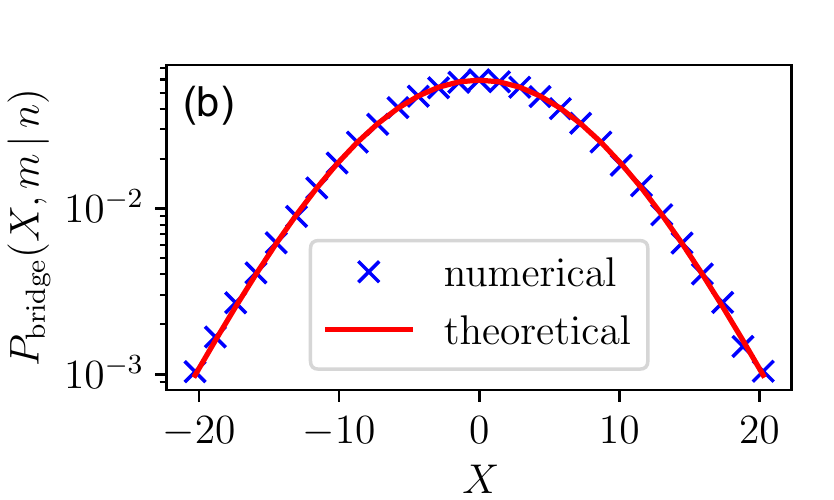}%
}\hfill
\caption{\textbf{Left panel (a):} Typical trajectories of a Student bridge random walk of $n=100$ steps generated using the ARS method on the effective distribution in Eq.~(\ref{eq:effS}). \textbf{Right panel (b):} Position distribution at $m=75$ for a Student bridge of $n=100$ steps generated using ARS on the the effective distribution. The position distribution $P_{\rm bridge}(X,m|n)$ obtained numerically by sampling $10^5$ trajectories using the ARS on the effective  distribution in Eq.~(\ref{eq:effS}) is compared with the theoretical prediction in Eq.~(\ref{PbridgeS}). }\label{fig:bridgeS}
\end{figure}

Our final example is a bridge random walk of $n$ steps where the free jump distribution is given by a Student's t-distribution (of parameter $3$) \cite{student}
\begin{align}
  f(\eta) = \frac{6 \sqrt{3}}{\pi  \left(\eta^2+3\right)^2}\,.\label{eq:stud2D}
\end{align} 
The Fourier transform of this jump distribution can be easily computed and is given by
\bea \label{Fourier_st}
\hat f(k)= e^{-\sqrt{3}\,|k|}(\sqrt{3} |k|+1) \;.
\eea
The backward propagator, using Eq. (\ref{eq:QI}), then reads 
\begin{align}
   Q(Y,m) = \int_{-\infty}^\infty \frac{dk}{2 \pi}\,e^{-m\sqrt{3} |k|} \left(\sqrt{3} |k|+1\right)^m\,e^{-ikY}\,,\label{eq:QST}
\end{align}
While this integral is hard to compute explicitly, it can still be evaluated numerically using the fast Fourier transform~\cite{fft}. The effective step distribution at the $m^\text{th}$ step (\ref{eq:eff}) is therefore given by 
\begin{align}
   \tilde f(\eta\,|\,Y,m,n) = \frac{\int_{-\infty}^\infty dk\,e^{-(n-m-1)\sqrt{3} |k|} \left(\sqrt{3} |k|+1\right)^{n-m-1}\,e^{-ik(Y+\eta)}}{\int_{-\infty}^\infty dk\,e^{-(n-m)\sqrt{3} |k|} \left(\sqrt{3} |k|+1\right)^{n-m}\,e^{-ikY}}\,  \frac{6 \sqrt{3}}{\pi  \left(\eta^2+3\right)^2}\,. \label{eq:effS}
\end{align}
Thus in this case, the effective jump distribution in Eq.~(\ref{eq:effS}) is again not explicit and it is therefore not easy
to draw a random number from this distribution. This is another example where the ARS method
discussed in Sec.~\ref{sec:ars} becomes handy. As in the previous examples, the propagator
$Q(Y,m)$ is also peaked around $Y=0$ and hence, in Eq.~(\ref{eq:eff}) one can use the inequality $Q(Y+\eta,n-m-1) \leq Q(0,n-m-1)$.
Hence, using this inequality in Eq.~(\ref{eq:eff}), we see from Eq.~(\ref{eq:ineq}) that a natural choice for $c_{m,n}(Y) \geq 1$ is
\begin{align}
 c_{m,n}(Y)  = \frac{Q(0,n-m-1)}{Q(Y,n-m)} = \frac{\int_{-\infty}^\infty dk\,e^{-(n-m-1)\sqrt{3} |k|} \left(\sqrt{3} |k|+1\right)^{n-m-1}}{\int_{-\infty}^\infty dk\,e^{-(n-m)\sqrt{3} |k|} \left(\sqrt{3} |k|+1\right)^{n-m}\,e^{-ikY}}\,.\label{eq:stud2cmn}
\end{align}
This yields the following acceptance probability, using Eq.~(\ref{eq:paccept}),
\begin{align}
    p_\text{accept}(\eta,Y,m,n) &= \frac{\int_{-\infty}^\infty dk\,e^{-(n-m-1)\sqrt{3} |k|} \left(\sqrt{3} |k|+1\right)^{n-m-1}\,e^{-ik(Y+\eta)}}{\int_{-\infty}^\infty dk\,e^{-(n-m-1)\sqrt{3} |k|} \left(\sqrt{3} |k|+1\right)^{n-m-1}}\,.\label{eq:pacceptS}
\end{align}
Using the ARS method discussed in Sec.~\ref{sec:bridgeRW}, we then generate Student bridge trajectories (see left panel in Fig.~\ref{fig:bridgeS}). In the right panel in Fig.~\ref{fig:bridgeS}, we computed numerically the probability distribution of the position at some intermediate time for a bridge random walk of $n=100$ steps, by sampling the jump distribution from Eq.~(\ref{eq:effS}) using the ARS method with the acceptance probability in (\ref{eq:pacceptS}). This is compared to the theoretical position distribution for the bridge which can be easily computed by substituting the free propagators $P(X,m) = \int_{-\infty}^\infty \frac{dk}{2 \pi}\,e^{-m\sqrt{3} |k|} \left(\sqrt{3} |k|+1\right)^m\,e^{-ikX}$ and $Q(X,m)$ from Eq.~(\ref{eq:QST}) in Eq.~(\ref{eq:bridgePQ}). This gives the theoretical position distribution 
\bea \label{PbridgeS}
P_{\rm bridge}(X,m|n) = \frac{\left[ \int_{-\infty}^\infty dk\,e^{-m\sqrt{3} |k|} \left(\sqrt{3} |k|+1\right)^m\,e^{-ikX}\right]\,\times  \left[\int_{-\infty}^\infty dk\, e^{-(n-m)\sqrt{3} |k|} \left(\sqrt{3} |k|+1\right)^{n-m}\,e^{-ikX}\right]}{2\pi\, \int_{-\infty}^\infty dk\,e^{-n\sqrt{3} |k|} \left(\sqrt{3} |k|+1\right)^n}\;.
\eea
 As can be seen in Fig.~\ref{fig:bridgeS} (right panel), the agreement is perfect.  

\section{Generalization to other constrained discrete-time random walks}
\label{sec:gen}
In Sec.~\ref{sec:bridgeRW}, we derived an effective step distribution for a bridge random walk. In this section, we generalize this construction to other discrete-time random walks,
such as the ``generalized bridge'', the random walk excursion and the random walk meander.

\subsection{Generalized bridge} For a generalized bridge, the random walk is constrained to start at the origin and finish at the final position $X_f$, not necessarily located at the origin. The random walk is described by the equation of motion (\ref{eq:XmI}) along with the constraint
\begin{subequations}
\begin{align}
  X_0 &= 0\,,\\
  X_n &= X_f\,.
\end{align}
\label{eq:genbridgeC}
\end{subequations}
\begin{figure}[ht]
  \begin{center}
    \includegraphics[width=0.5\textwidth]{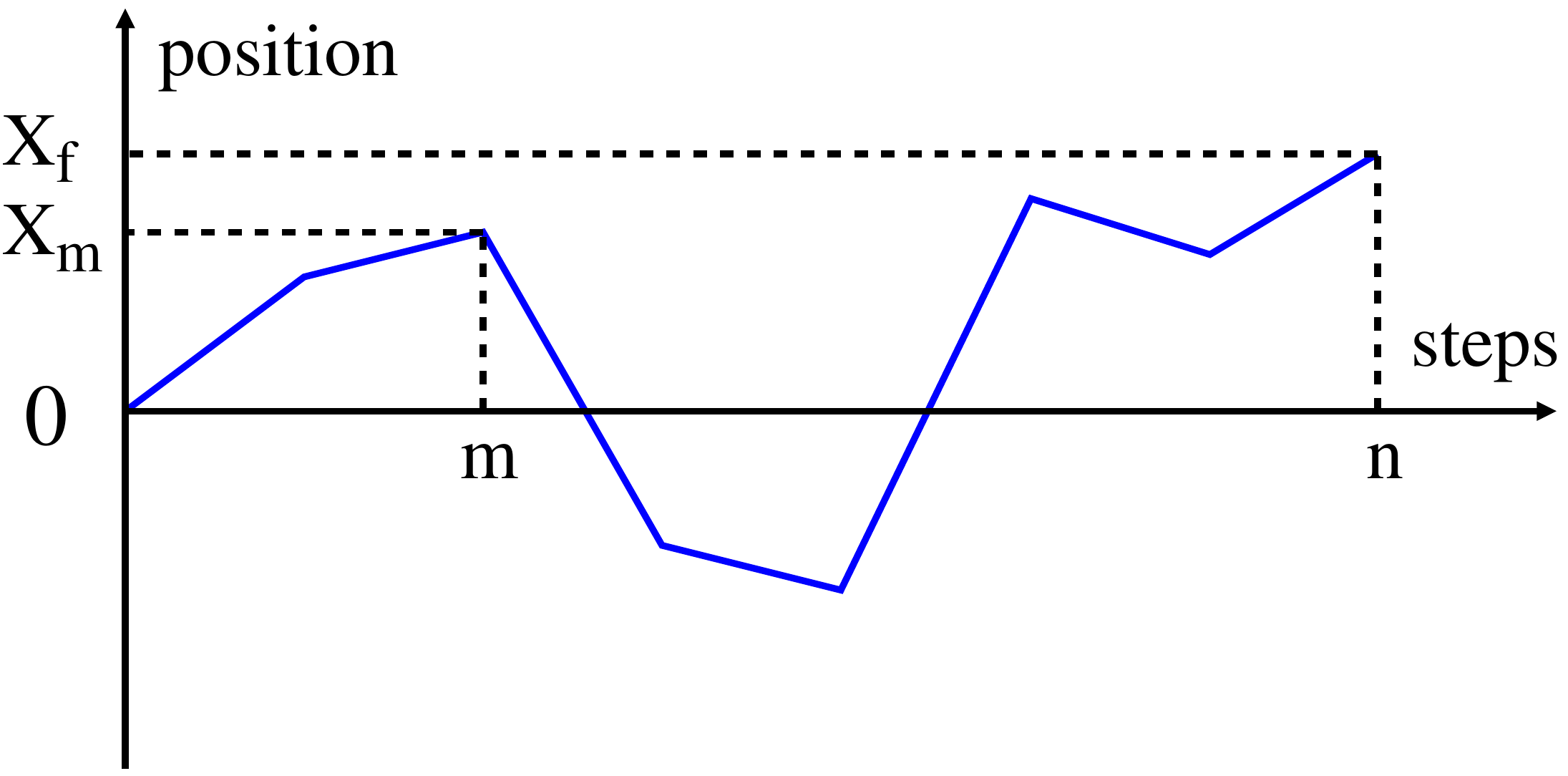}
    \caption{A generalized bridge random walk of $n$ steps is a random walk that is constrained to start at the origin and finish at $X_f$ after $n$ steps. Due to the Markov property, a generalized bridge random walk can be decomposed into two independent parts: a left part over the interval $[0,m]$, where it propagates from the $0$ to $X_m$ and a right part over the interval $[m,n]$, where it propagates goes from $X_m$ to $X_f$.}
    \label{fig:bridgegen}
  \end{center}
\end{figure}
An analogous reasoning as in Sec.~\ref{sec:bridgeRW} gives the generalization of Eq.~(\ref{eq:bridgePQ}) for the propagator (see Fig.~\ref{fig:bridgegen})
\begin{align}
  P_\text{generalized bridge}(X,m\,|\,n) = \frac{P(X,m)\,Q(X-X_f,n-m)}{P(X=X_f,n)}\,,\label{eq:bridgePQg}
\end{align}
where $P(X,m)$ and $Q(X,m)$ are the forward and backward free propagators respectively defined in Eq.~(\ref{eq:P}) and Eq.~(\ref{eq:Q}). The effective step distribution (\ref{eq:eff}) straightforwardly generalizes to 
\begin{align}
  \tilde f(\eta\,|\,X,X_f,m,n) = f(\eta) \,\frac{Q(X-X_f+\eta,n-m-1)}{Q(X-X_f,n-m)}\,, \label{eq:effGe}
\end{align}
where, as before, $f(\eta)$ is the free jump distribution and $\tilde f(\eta\,|\,X,X_f,m,n)$ is the effective jump distribution at time $m$ given the position $X$
at time $m$ and the final position $X_f$ at time $n$. Following the analysis done for the standard bridge in Sec.~\ref{sec:examples} for different jump distributions $f(\eta)$,
one can also obtain explicit results for different examples but we do not repeat the details here. We just mention only one example, namely the Gaussian jump distribution $f(\eta) = e^{-\eta^2/(2 \sigma^2)}/\sqrt{2 \pi \sigma^2}$.
Here, following the steps in Sec.~\ref{sec:Bg}, one can compute explicitly the effective jump distribution in Eq.~(\ref{eq:effGe}). It turns out to be Gaussian again with a nonzero mean 
\begin{align}
   \tilde f (\eta|X,X_f,m,n)=\frac{1}{\sqrt{2\pi\,\sigma_{m,n}^2}}\,e^{-\frac{(\eta-\mu_{m,n})^2}{2\sigma_{m,n}^2}}\,,\label{eq:effGeg}
  \end{align}
  where
  \begin{align}
   \mu_{m,n}&=-\frac{X-X_f}{n-m}\,,\label{eq:mumng}\\
   \sigma_{m,n}^2&=\frac{(n-m-1)\,\sigma^2}{n-m}\,,\label{eq:sigmamng}
\end{align}
which recovers the result in Eq.~(\ref{eq:effG}) for $X_f=0$. In this case, the discrete Langevin equation analogous to Eq.~(\ref{eq:eomG}) reads
\begin{align}
  X_{m+1} = X_m - \frac{X_m-X_f}{n-m} + \frac{\sqrt{n-m-1}}{\sqrt{n-m}}\,\sigma\,\eta_m \quad, \quad 0 \leq m < n \;. \label{eq:eomG2}
\end{align}
In the continuous time limit, it approaches the Langevin equation 
 \begin{align}
   \dot X(t) &= - \frac{X(t)-X_f}{t_f-t}\, + \sqrt{2\,D}\,\xi(t)\,,\label{eq:contTg}
\end{align}
derived in Ref.~\cite{MajumdarEff15}.

\subsection{Excursion}
\begin{figure}[htb]
\subfloat{%
 \includegraphics[width=0.5\textwidth]{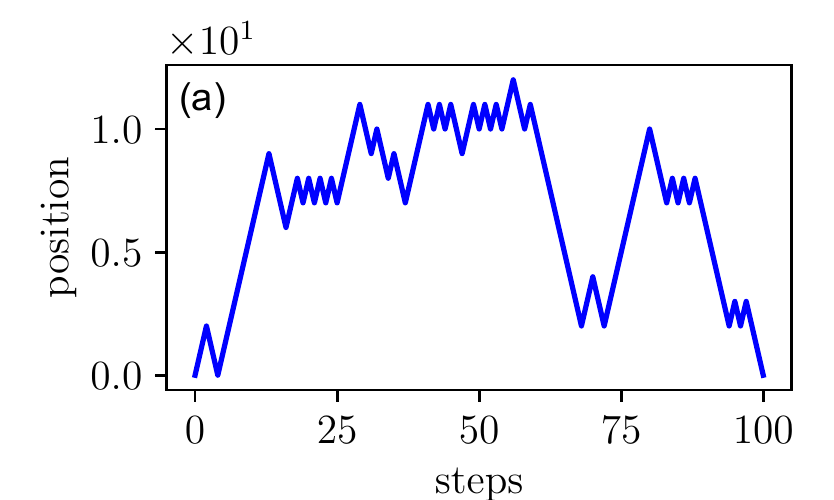}%
}\hfill
\subfloat{%
  \includegraphics[width=0.5\textwidth]{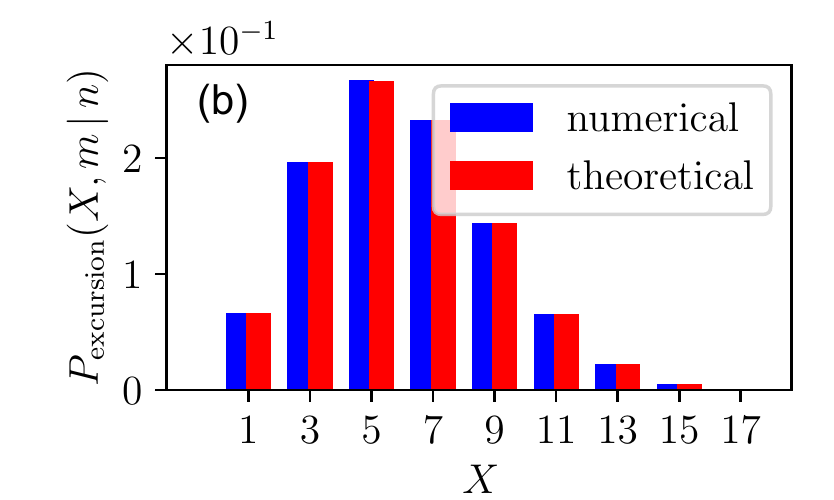}%
}\hfill
\caption{\textbf{Left panel (a):} A typical trajectory of a lattice excursion random walk of $n=100$ steps generated by the effective jump distribution in Eq.~(\ref{eq:feffex}). \textbf{Right panel (b):} Position distribution at $m=75$ for a lattice excursion of $n=100$ steps. The position distribution $P_{\rm excursion}(X,m|n)$ obtained numerically by sampling $10^{5}$ trajectories using the jumps from the effective  distribution (\ref{eq:feffex}) is compared with the theoretical prediction in Eq.~(\ref{Pexclat}). }\label{fig:excS}
\end{figure}

An excursion is a bridge random walk that is constrained to stay above the origin. It is described by the equation of motion (\ref{eq:XmI}) along with the constraints
\begin{subequations}
\begin{align}
  X_0 &= 0\,,\\
  X_n &= 0\,,\\
  X_m &\geq  0 \,,\quad m = 1,\ldots, n-1\,.
\end{align}
\label{eq:excC}
\end{subequations}
\begin{figure}[ht]
  \begin{center}
    \includegraphics[width=0.5\textwidth]{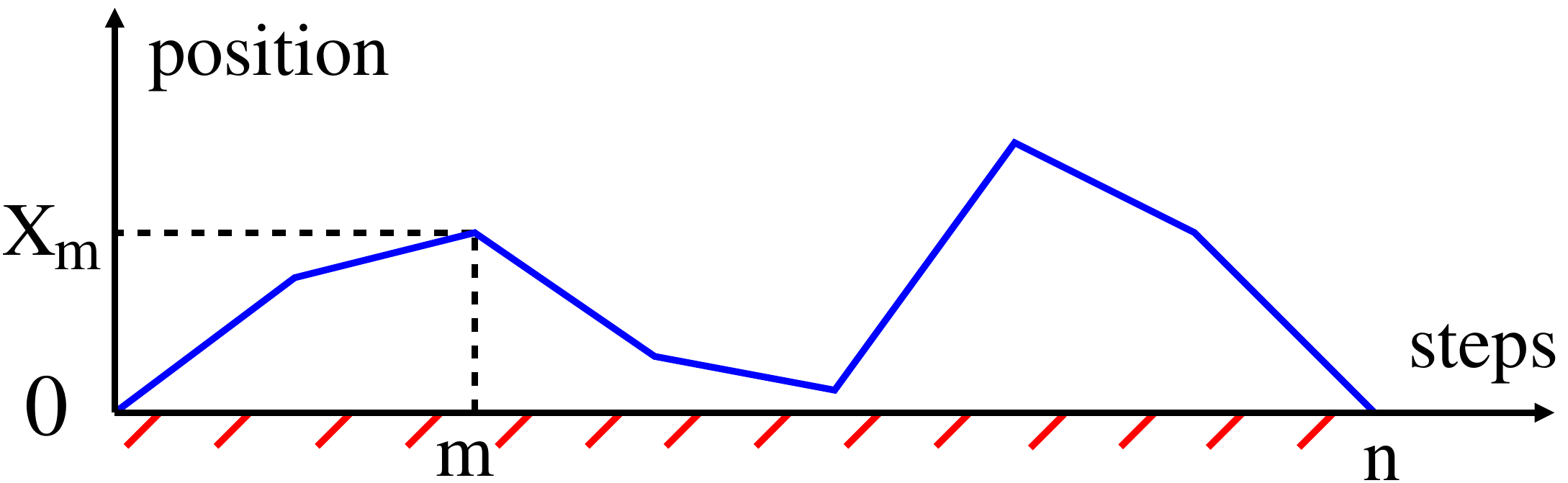}
    \caption{An excursion random walk of $n$ steps is a bridge random walk that is constrained to remain above the origin (the absorbing origin at $X=0$ is indicated by the red dashes on the horizontal axis). Due to the Markov property, an excursion random walk can be decomposed into two independent parts: a left part over the interval $[0,m]$, where it propagates from the $0$ to $X_m$ while staying positive and a right part over the interval $[m,n]$, where it propagates goes from $X_m$ to $0$ while staying positive.}
    \label{fig:excursion}
  \end{center}
\end{figure}
The propagator of an excursion becomes (see Fig.~\ref{fig:excursion})
\begin{align}
  P_\text{excursion}(X,m\,|\,n) = \frac{P_\text{absorbing}(X,m)\,Q_\text{absorbing}(X,n-m)}{P_\text{absorbing}(X=0,n)}\,,\label{eq:bridgePQe}
\end{align}
where $P_\text{absorbing}(X,m)$ and $Q_\text{absorbing}(X,m)$ are the forward and backward propagators in the presence of an absorbing boundary located at the origin. They satisfy the following recursion relations 
\begin{subequations}
\begin{align}
  P_\text{absorbing}(X,m) &= \int_{0}^\infty dY P_\text{absorbing}(Y,m-1)\,f(X-Y)\,,\label{eq:Pa}\\
    Q_\text{absorbing}(X,m) &= \int_{0}^\infty dY Q_\text{absorbing}(Y,m-1)\,f(Y-X)\,,\label{eq:Qa}
\end{align}
\label{eq:PQa}
\end{subequations}
where $P_\text{absorbing}(X,0)=\delta(X)$ and $Q_\text{absorbing}(X,0)=\delta(X)$. Note that the integral equations in Eq.~(\ref{eq:PQa}) are of the Wiener-Hopf type (where the integrals extends only over the semi-infinite line) and are notoriously difficult to solve explicitly for arbitrary jump distribution $f(\eta)$ \cite{Leuven}. Following the steps in Sec.~\ref{subsec1}, the analogue of Eq.~(\ref{eq:eff}) for the effective distribution now becomes 
\begin{align}
  \tilde f(\eta\,|\,Y,m,n) = f(\eta) \,\frac{Q_\text{absorbing}(Y+\eta,n-m-1)}{Q_\text{absorbing}(Y,n-m)}\,,\label{eq:effGee}
\end{align}
where $f(\eta)$ is the free jump distribution and $Q_\text{absorbing}(X,m)$ is the solution of the Wiener-Hopf equation (\ref{eq:Qa}). In the absence of an explicit solution for this Wiener-Hopf equations, it is then hard to compute the effective jump distribution $\tilde f$ for general $f(\eta)$. There is however one exactly solvable case. This corresponds to lattice random 
walk with jumps $\eta= \pm 1$. In this case, the lattice excursion trajectories are called Dyck paths, which have been studied extensively \cite{flajolet,louchard}. In this case the
propagator $Q_\text{absorbing}(Y,m)$ can be computed explicitly using the method of images and one gets (see e.g. \cite{wergen})  
\begin{align}
Q_\text{absorbing}(Y,m) = 2^{-m}
   \binom{m}{\frac{m+Y}{2}}-2^{-
   m} \binom{m}{\frac{m+Y+2}{2}
   }\,,\label{eq:Qabssim}
\end{align}
where we recall that $(m+Y)$ is even. In this case, the effective jump distribution in Eq.~(\ref{eq:effGee}) can be computed explicitly and we get
\begin{align}
   \tilde f(\eta\,|\,Y,m,n) = \frac{1}{2}\left(1+\frac{n-m-Y (Y+2)}{ (Y+1)
   (n-m)}\right) \delta(\eta-1)
   + \frac{1}{2}\left(1-\frac{n-m-Y (Y+2)}{ (Y+1)
   (n-m)}\right)\delta(\eta+1)\,,\label{eq:feffex}
\end{align}
where, again, we have the condition that $Y+n-m$ is even. We note that this result was result in Eq. (\ref{eq:feffex}) was also derived recently in Ref. \cite{Rose21} using the approach of generalized Doob transform. In this special case, the ARS method is not needed and we can directly sample the jumps from the effective jump distribution in Eq.~(\ref{eq:feffex}) to obtain excursion trajectories (see left panel in Fig.~\ref{fig:excS}). In the right panel in Fig.~\ref{fig:excS}, we computed numerically the probability distribution of the position at some intermediate time by sampling the jump distribution from Eq.~(\ref{eq:feffex}). This is compared to the theoretical position distribution for the excursion which can be easily computed by substituting the propagators $P_\text{absorbing}(X,m) = 2^{-m}
   \binom{m}{\frac{m+X}{2}}-2^{-
   m} \binom{m}{\frac{m+X+2}{2}
   }$ and $Q_\text{absorbing}(X,m)$ from Eq.~(\ref{eq:Qabssim}) in Eq.~(\ref{eq:bridgePQe}), which gives
\bea \label{Pexclat}
 P_\text{excursion}(X,m\,|\,n)  =   \frac{\left[\binom{m}{\frac{m+X}{2}}-\binom{m}{\frac{m+X+2}{2}}\right] \left[\binom{n-m}{\frac{n-m+X}{2}}-\binom{n-m}{\frac{n-m+X+2}{2}}\right]}{\binom{n}{\frac{n}{2}}-\binom{n}{\frac{n+2}{2}}}\;,
\eea
which is nonzero only if $(m+X)$ as well as $n$ are both even numbers.  As can be seen in Fig.~\ref{fig:excS} (right panel), the agreement is perfect.  

\subsection{Meander}
A meander is a random walk that is constrained to stay above the origin. It is described by the equation of motion (\ref{eq:XmI}) along with the constraints
\begin{subequations}
\begin{align}
  X_0 &= 0\,,\\
  X_m &\geq  0 \,,\quad m = 1,\ldots, n\,.
\end{align}
\label{eq:meaM}
\end{subequations}
\begin{figure}[ht]
  \begin{center}
    \includegraphics[width=0.5\textwidth]{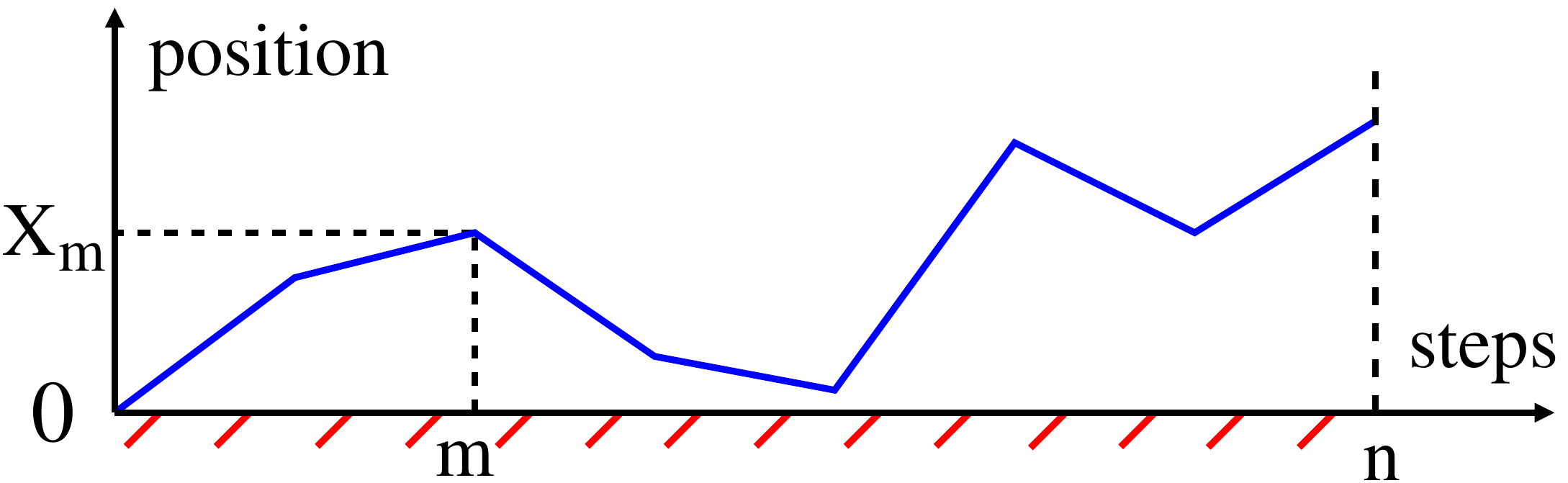}
    \caption{A meander random walk of $n$ steps is a random walk that is constrained to remain above the origin. Due to the Markov property, a meander random walk can be decomposed into two independent parts: a left part over the interval $[0,m]$, where it propagates from the $0$ to $X_m$ while staying positive and a right part over the interval $[m,n]$, where it propagates to any position while staying positive. The horizontal line with red dashes just indicates that the origin at $X=0$ is absorbing.}
    \label{fig:meander}
  \end{center}
\end{figure}
The propagator of a meander becomes (see Fig.~\ref{fig:meander})
\begin{align}
  P_\text{meander}(X,m\,|\,n) = \frac{P_\text{absorbing}(X,m)\,S(X,n-m)}{S(0,n)}\,,\label{eq:bridgePQm} \;.
\end{align}
Here, $P_\text{absorbing}(X,m)$ is the forward propagator in the presence of an absorbing boundary located at the origin defined in Eq.~(\ref{eq:Pa}). The quantity 
$S(X,m)$ denotes the survival probability, i.e., the probability that the walker does not cross the origin during $m$ steps given that it started at $X$. This equation is again understood 
by splitting a typical trajectory into two parts as in Fig. \ref{fig:meander}. On the left part, the trajectory goes from the origin to $X$ after $m$ steps, while staying nonnegative -- this occurs with probability $P_\text{absorbing}(X,m)$. On the right side, the trajectory starts from $X$ and stays nonnegative up to $n-m$ steps -- the probability for this event is the survival probability $S(X,n-m)$. This equation is normalized by the number of all trajectories that start at the origin and stay nonnegative up to step $n$. This is proportional to $S(0,n)$. The numerator on the rhs in Eq. (\ref{eq:bridgePQm}), when integrated over $X \in [0, +\infty)$ indeed gives $S(0,n)$, which shows that $P_\text{meander}(X,m\,|\,n)$ is normalised to unity.

\begin{figure}[htb]
\subfloat{%
 \includegraphics[width=0.5\textwidth]{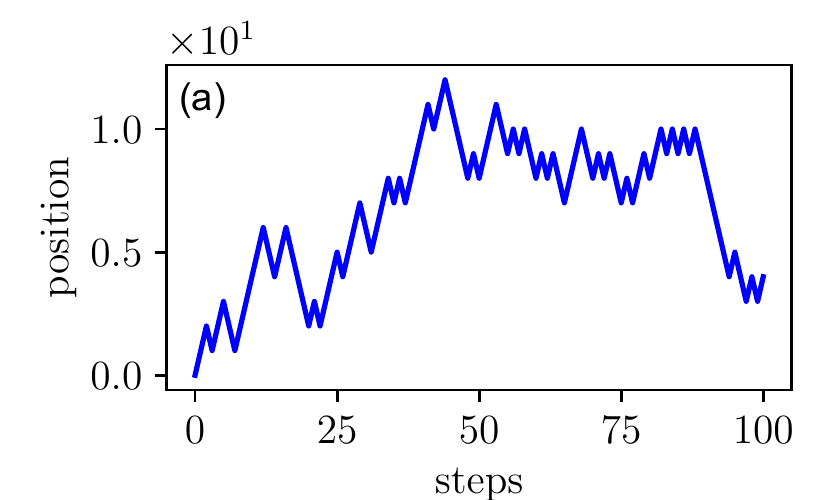}%
}\hfill
\subfloat{%
  \includegraphics[width=0.5\textwidth]{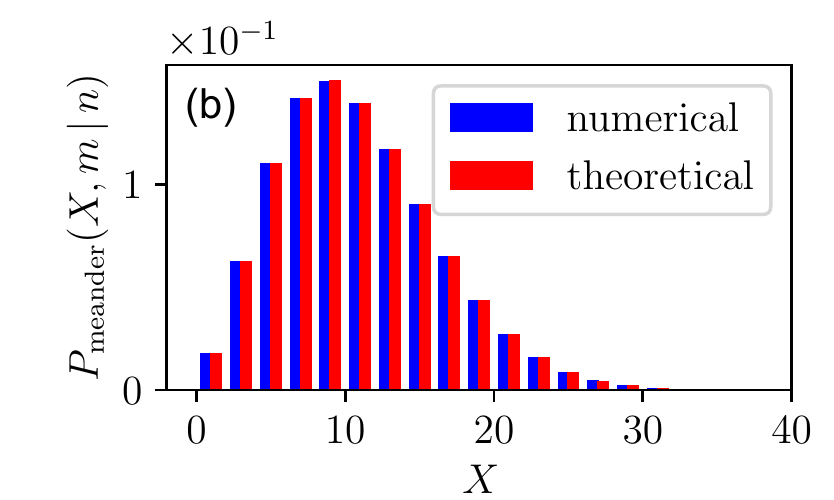}%
}\hfill
\caption{\textbf{Left panel (a):} A typical trajectory of a lattice meander random walk of $n=100$ steps generated by the effective jump distribution in Eq.~(\ref{eq:effMea}). \textbf{Right panel (b):} Position distribution at $m=75$ for a lattice meander of $n=100$ steps. The position distribution $P_{\rm meander}(X,m|n)$ obtained numerically by sampling $10^{7}$ trajectories using the jumps from the effective  distribution (\ref{eq:effMea}) is compared with the theoretical prediction in Eq.~(\ref{Pmealat}). }\label{fig:meaS}
\end{figure}

The survival probability $S(X,m)$ satisfies the recursion relation
\begin{align}
S(X,m) &= \int_{0}^\infty dY S(Y,m-1)\,f(Y-X)\,,
\label{eq:S}
\end{align}
where $S(X,0)=\Theta(X)$ and $\Theta(z)$ is the Heavyside step function such that $\Theta(z)=1$ if $z\geq 0$ and $\Theta(z)=0$ otherwise. Eq. (\ref{eq:S}) is easy to understand: consider a trajectory of $m$ steps, starting at $X$ and staying nonnegative up to step $m$. Suppose in the first step the walker moves from $X$ to $Y \geq 0$, which occurs with a probability $f(Y-X)\, dY$ and then,  starting at the new initial position $Y$ the walker stays positive up to $m-1$ steps. The probability for this latter event is just $S(Y,m-1)$. Integrating over all possible values of $Y \in [0, +\infty)$ gives Eq. (\ref{eq:S}). The integral equation (\ref{eq:S}) is of the Wiener-Hopf type and is therefore difficult to solve explicitly for arbitrary jump distribution $f(\eta)$ \cite{Leuven}.  Following the steps in Sec.~\ref{subsec1}, the analogue of Eq.~(\ref{eq:eff}) for the effective distribution now becomes 
\begin{align}
   \tilde f(\eta\,|\,X,m,n) = f(\eta) \,\frac{S(X+\eta,n-m-1)}{S(X,n-m)}\,, \label{eq:effGm}
\end{align}
where $f(\eta)$ is the free jump distribution and $S(X,m)$ is the solution of the Wiener-Hopf equation (\ref{eq:S}). In the absence of an explicit solution for this Wiener-Hopf equations, it is then hard to compute the effective jump distribution $\tilde f$ for general $f(\eta)$. As in the previous section, there is one exactly solvable case which is the lattice random walk with jumps $\eta= \pm 1$. The survival probability $S(Y,m)$ can be obtained by summing the backward propagator over all the possible final positions $X_f$ (see e.g. \cite{wergen}):
\begin{align}
S(Y,m) = \sum_{X_f=0}^{Y+m} 2^{-m}
   \binom{m}{\frac{m+X_f-Y}{2}}-2^{-
   m} \binom{m}{\frac{m+X_f+Y+2}{2}
   }\,,\label{eq:Ssim}
\end{align}
where the summand is nonzero only when $(m+X_f-Y)$ is even. In this case, the effective jump distribution in Eq.~(\ref{eq:effGm}) becomes
\begin{align}
  \tilde f(\eta\,|\,X,m,n) =  \,\frac{S(X+1,n-m-1)}{2\,S(X,n-m)}\,\delta(\eta-1) + \frac{S(X-1,n-m-1)}{2\,S(X,n-m)}\,\delta(\eta+1) \,,\label{eq:effMea}
\end{align}
where $S(X,m)$ is given in Eq.~(\ref{eq:Ssim}). Even if there is no explicit expression for the sum in Eq.~(\ref{eq:Ssim}), it can still be evaluated numerically straightforwardly. As in the previous section, the ARS method is not needed and we can directly sample the jumps from the effective jump distribution in Eq.~(\ref{eq:effMea}) to obtain meander trajectories (see left panel in Fig.~\ref{fig:meaS}). In the right panel in Fig.~\ref{fig:meaS}, we computed numerically the probability distribution of the position at some intermediate time by sampling the jump distribution from Eq.~(\ref{eq:effMea}). This is compared to the theoretical position distribution for the meander which can be easily computed by substituting the propagator $P_\text{absorbing}(X,m) = 2^{-m}
   \binom{m}{\frac{m+X}{2}}-2^{-
   m} \binom{m}{\frac{m+X+2}{2}
   }$ and $S(X,m)$ from Eq.~(\ref{eq:Ssim}) in Eq.~(\ref{eq:bridgePQm}), which gives
\bea \label{Pmealat}
 P_\text{meander}(X,m\,|\,n)  =    \frac{\left[\binom{m}{\frac{m+X}{2}}-\binom{m}{\frac{m+X+2}{2}}\right] \left[\sum_{X_f=0}^{X+n-m} 
   \binom{n-m}{\frac{n-m+X_f-X}{2}}- \binom{n-m}{\frac{n-m+X_f+X+2}{2}
   }\right]}{\binom{n}{n-\left\lfloor \frac{n}{2}\right\rfloor }}\;,
\eea
where we used that $S(0,n)=2^{-n}\binom{n}{n-\left\lfloor \frac{n}{2}\right\rfloor }$ where $\left\lfloor \frac{n}{2}\right\rfloor$ denotes the integer part of $n/2$. Note that the summand in Eq. (\ref{Pmealat}) is nonzero only when $(n-m+X_f-X)$ is even. As can be seen in Fig.~\ref{fig:meaS} (right panel), the agreement is perfect.  

\section{Summary and outlook}
\label{sec:ccl}

In this paper, we have studied discrete-time random walk bridges, where the random walk starts at the origin and is constrained to return to the origin after a fixed number $n$ of steps. One of the challenges was to find an effective jump distribution that is local in time and yet
takes into account automatically the global bridge constraint, and is valid for arbitrary bare jump distributions. We have derived an exact formula for this effective jump distribution and computed it explicitly in a few examples of bare jump distributions. Furthermore, our method provides, for arbitrary jump distributions, an exact and efficient numerical algorithm to generate random walk bridge trajectories with the correct statistical weight. We have also provided a numerical method based on acceptance-rejection sampling which is versatile and powerful to generate a random jump from the effective distribution, even when the effective distribution has a complicated functional form. Going beyond the simple bridges, 
we have further extended our method to other constrained discrete-time random walks, such as ``generalized bridges'' (where the endpoint is different from the starting point), excursions and meanders. 

One interesting application of our method is in the context of extreme value statistics for constrained discrete-time random walks \cite{review_EVS}. For such walks, there have been a lot of interesting analytical results that have been derived recently for arbitrary jump distribution, an example being the expected maximum of a random walk bridge of $n$ steps \cite{FrankeSur12,Coffman,comtet,mounaix,grebenkov}. Another example is the exact distribution of the maximal relative height of a one-dimensional discrete solid-on-solid model in the stationary state 
with periodic boundary condition \cite{Schehr06SOS}.  In order to verify such analytical predictions numerically, one needs to generates efficiently the discrete-time bridge trajectories with the correct statistical weight. The method presented in this paper will be useful for this purpose. In a recent work \cite{Rose21}, the generalized Doob's transform has been used to develop a reinforced learning approach to generate rare
atypical trajectories, with a given statistical weight -- we hope that the method developed in this paper will also be useful in such applications.   
 
In this paper, we focused on constrained discrete-time random walks in one-dimension. It would be interesting to generalize our results for the effective jump distribution for discrete-time
bridges in higher dimensions. We note that discrete-time random walks in $d$-dimensions is different from $d$ independent one-dimensional discrete-time random walks (a property which however holds for continuous-time Brownian motion). Even for $d$-dimensional lattice random walk bridges, it would be interesting to compute the effective jump distribution from a given site at a fixed intermediate time. 

\appendix
\section{Correlator of the Gaussian bridge}
\label{app:gbridge}
The correlator of the generated bridge in Eq.~(\ref{bcbm_discr}) is
\begin{align}
  \langle X^\text{bridge}_i X^\text{bridge}_j  \rangle = \langle X^\text{free}_i\,X^\text{free}_j \rangle - \frac{i}{n}\,\langle X^\text{free}_j\,X^\text{free}_n \rangle- \frac{j}{n}\,\langle X^\text{free}_i\,X^\text{free}_n \rangle+ \frac{i\,j}{n^2}\,\langle [X^\text{free}_n]^2\rangle\,.\label{eq:corrG}
\end{align}
Given that the correlator of the free Gaussian random walk is
\begin{align}
  \langle X^\text{free}_i\,X^\text{free}_j \rangle = \sigma^2\,\min(i,j)\,,\label{eq:corrGF}
\end{align}
we find
\begin{align}
  \langle X^\text{bridge}_i X^\text{bridge}_j  \rangle = \sigma^2\left[\min(i,j) - \frac{i\,j}{n}\right]\,,\label{eq:corrGB}
\end{align}
which indeed corresponds to the bridge correlator (which one can compute from the propagator in Eq.~(\ref{eq:backPropG})).

\begin{acknowledgments}
BD wishes to thank F. Mori for an introduction to the acceptance-rejection sampling method. We thank J.~P.~Garrahan for useful comments and discussions. This work was partially supported by the Luxembourg National Research Fund (FNR) (App. ID 14548297).
\end{acknowledgments}

\end{document}